\documentclass[%
 reprint,
 amsmath,amssymb,
 aps,
pra,
]{revtex4-2}

\usepackage{graphicx}% Include figure files
\usepackage{dcolumn}% Align table columns on decimal point
\usepackage{bm}% bold math
\usepackage{color}
\usepackage{xcolor}

\begin{document}

%\preprint{APS/123-QED}

\title{Control of spatial four wave mixing efficiency in Bessel beams using longitudinal intensity shaping}% Force line breaks with \\

\author{Ismail Ouadghiri-Idrissi, John M. Dudley}
\author{Francois Courvoisier}%
\email{francois.courvoisier@femto-st.fr}
\affiliation{%
 Institut FEMTO-ST, UMR 6174 CNRS University Bourgogne Franche-Comt\'{e},\\
 15B Avenue des Montboucons, F-25030 Besan\c con, France
}%

\date{\today}% It is always \today, today,

\begin{abstract}
Diffraction-free Bessel beams have attracted major interest because of their stability even in regimes of nonlinear propagation and filamentation. However, Kerr nonlinear couplings are known to induce significant longitudinal intensity modulation, detrimental to the generation of uniform plasma or for applications in the processing of transparent materials. These nonlinear instabilities arise from the generation of new spatio-spectral components through an initial stage of continuous spectral broadening  followed by four wave mixing.  In this paper, we investigate analytically and numerically these processes and show that nonlinear instabilities can be controlled through shaping the spatial spectral phase of the input beam. This opens new routes for suppressing the nonlinear growth of new frequencies and controlling ultrashort pulse propagation in dielectrics.
\end{abstract}

%\keywords{Suggested keywords}%Use showkeys class option if keyword
                              %display desired
\maketitle

%\tableofcontents

\section{Introduction}

Diffraction-free Bessel beams are formed from a conical energy flow and yield a near-uniform intensity distribution along a line focus \cite{Durnin1987}. For high power laser pulses injected into transparent dielectrics, this beam structure can sustain quasi propagation-invariant regimes of filamentation, which are highly advantageous in reducing nonlinear distortion and instabilities during propagation. This has been shown to yield significant improvement in controlling energy deposition, and has been the subject of intense interest for generating long and uniform plasma channels \cite{Porras2004,Polesana2008,Dota2012} and creating high aspect ratio structures in micro- and nano-machining applications \cite{Bhuyan2010,Bhuyan2011,Garzillo2016,Courvoisier2016,Stoian2018}. The application of Bessel beams in laser processing and filamentation has also been demonstrated for Bessel vortices \cite{Xie2015}.

However, although the level of nonlinear distortion during Bessel beam propagation is greatly reduced compared to Gaussian beams at comparable peak intensities, residual nonlinear instability effects can still occur and induce significant oscillations of the on-axis intensity \cite{Gadonas2001,Polesana2008,Andreev1991}. This is clearly detrimental for the creation of longitudinally-uniform structures in material processing, and thus controlling these instabilities is of central importance to extend the applicability of Bessel beams. 

Such nonlinear instabilities were first investigated in 1989 in association with the optical breakdown in gases and discussed in terms of cubic nonlinearity and plasma dynamics \cite{Andreev1991}. Gadonas {\it et al.} subsequently investigated the relation between nonlinear instabilities in Bessel beams and the distortion of their spatial spectrum \cite{Gadonas2001}. Using phase-matching arguments and considering four wave mixing interactions, they showed that a Bessel beam with radial wavevector $k_{r0}$ can sustain Kerr self-action which deforms its spatial spectrum to generate two additional spectral components: an axial wave component with $k_r=0$ and a secondary conical wave with $k_r=\sqrt{2}k_{r0}$ \cite{Gadonas2001,Pyragaite2006}. Their numerical simulations highlighted the development of an intensity modulation of Bessel beams along the propagation direction which was explained by the interference of the conical beam with the axial wave component. Experimental signatures of this spectrum distortion was demonstrated in \cite{Gadonas2001,Pyragaite2006,Polesana2007}. 

In order to achieve stable propagation of Bessel beams, it is necessary to overcome these Kerr-induced instabilities \cite{Johannisson2003}. Porras {\it et al} investigated the stability of Bessel beams in the presence of nonlinear losses and showed that nonlinear instabilities can be significantly suppressed if multiphoton absorption prevails over Kerr nonlinearity \cite{Porras2004}. This condition can be satisfied for relatively high input power and high cone angles \cite{Porras2004,Couairon2013}, and experimental demonstration was performed in \cite{Porras2004,Polesana2007,Gaizauskas2006}. However, since this imposes strong constraints on the geometrical and physical properties of the input Bessel beam, this approach is not suitable in all applications.  In other work, Polesana {\it et al.} investigated the effect of the input injection condition to the nonlinear medium \cite{Polesana2008} and showed that the Kerr-induced instabilities can be attenuated if the Bessel beam is progressively formed inside the Kerr medium. In contrast, if the Bessel beam is formed prior entering the medium, it was shown to exhibit significant instabilities and quasi-periodic intensity modulation along the propagation direction. 
 
Recently, we have shown that these instabilities can be significantly mitigated by appropriate control of the intensity evolution of Bessel beams along the propagation distance \cite{Ouadghiri2017}. In particular, we qualitatively identified the role of the spatial spectral phase in reducing the efficiency of Four Wave Mixing (FWM) and the growth of related nonlinear instabilities. In this paper we present an extended analysis of this problem, and through both analysis of the FWM process and numerical simulations, we obtain significant new insights into the physical origins of these instabilities, and identify particular quantitative parameter regimes in which they can be suppressed. Note that our approach is complementary to the work in Refs \onlinecite{Porras2004,Polesana2008,Porras2015,Porras2016} where propagation is analyzed in terms of a linearized stability analysis, and where nonlinear losses are shown to give rise to an attractor (nonlinear unbalanced Bessel beam).  Here, in contrast, we neglect nonlinear losses to isolate the effect of input beam shaping on Four-wave mixing efficiency and growth of the intensity oscillations in the Bessel beam.

This paper is organized as follows.  We first develop in Section II an analytical model of four wave mixing that allows us to describe the cascade of nonlinear effects that lead to the generation of new {\it spatial} spectral components. In section III, we examine the dynamics of the growth of new spatial spectral components and show that an initial phase of spectral broadening is the major driver for instabilities. In section IV, we develop a reduced model with only the essential terms responsible for spectral broadening which allows us to take into account the spectral width - in the spatial domain - of Bessel beams. In section V, we will use this model to discuss the dependence of nonlinear instabilities on the precise injection condition of an incident Bessel beams with respect to a nonlinear medium (referred to as the ``soft'' and ``abrupt'' transition cases) considering in particular the influence of the initial spectral phase. Finally in section VI, we consider previous numerical results studying  the nonlinear propagation of Bessel beams with shaped on-axis intensity profiles \cite{Ouadghiri2017}, with our modelling allowing us to understand why specific input spectral phases lead to the reduction of nonlinear instabilities.

%%%%%%%%%%%%%%%%%%%%%%%%%%%%%%%%%%%%%%%%%%%%%%%%%%%%%%%%%%%%%%%%%%%%%%%%%%%%%%

\section{Four wave mixing in Bessel beams}

In 1996, Tewari {\it et al} introduced a theoretical model to describe  Third Harmonic Generation (THG) in Bessel beams \cite{Tewari1996}. Our theoretical model follows the same approach. However, instead of THG nonlinear terms, we consider four wave mixing (FWM) interactions to study nonlinear spectral distortions in Bessel beams related to Kerr-induced instabilities. Importantly, our focus is on the growth of spatial frequencies $k_r$ in a monochromatic model such that we do not consider dynamics related to the  growth of new temporal frequencies $\omega$.  However, we stress that the formalism developed here could be extended to describe the full picture of simultaneous nonlinear dynamics in both spatial and temporal domains.  We note the monochromatic approximation has proven successful in interpreting experimental results obtained with pulsed beams with typical pulse durations of some hundreds fs, peak powers in the order of TW.cm$^{-2}$, cone angles of a few degrees \cite{Porras2004,Polesana2007,Polesana2008,Xie2015,Gadonas2001} as we will use here as numerical examples, for the nonlinear propagation of ultrafast Bessel beams in fused silica. This approximation is valid as long as temporal reshaping is not too strong.

We first compute the third order nonlinear polarization and select the relevant nonlinear terms which have direct impact on the generation of new spectral components. We will then include these terms in the Helmholtz equation which we study analytically and numerically. 

\subsection{Third order nonlinear polarization \label{sec:ThirdOrderPolar}} 

Nonlinearity in Kerr media appears through the third order nonlinear polarization $\textbf{P}_{NL}$, written as:
  \begin{equation}
\textbf{P}_{NL}=\varepsilon_0\chi^{(3)}\textbf{E}\textbf{E}\textbf{E}
\label{P_NL}
 \end{equation} 
where $\varepsilon_0$ is the dielectric permittivity in vacuum, $\chi^{(3)}$ is the third order susceptibility and $ \textbf{E}$ is the total electric field. We consider monochromatic waves, oscillating at (temporal) frequency $\omega_0$. We consider four waves $\textbf{E}_j$, of amplitude $A_j$, linearly polarized along the same axis $\textbf{x}$ and possessing different longitudinal spatial frequencies $k_{zj}$. The total electric field is then written:
\begin{align}
\textbf{E} &= \dfrac{1}{2}\textbf{x} \sum \limits_{j=1}^4A_j \mathrm{exp}[i( \omega_0t-k_{zj}z )]\ +\ c.c 
\label{E_tot}
 \end{align}
If we substitute Eq. (\ref{E_tot}) in Eq. (\ref{P_NL}), we find that the nonlinear polarization consists of many terms covering all possible nonlinear interactions including terms responsible for Third Harmonic Generation (THG) and those preserving the same temporal frequency $\omega_0$. Neglecting THG terms because we investigate only the $\omega_0$ components, the nonlinear polarization can be written as: 

\begin{align}
\textbf{P}_{NL,\, \omega_0}= \dfrac{3}{8}\varepsilon_0\chi^{(3)}\textbf{x}\ \Big[ p_{NL,\, \omega_0}\mathrm{e}^{i\omega_0t} + c.c \Big] 
\label{p_NL2}
 \end{align} 

 where
 \begin{widetext}
\begin{align}
 p_{NL,\, \omega_0} = & \sum_{j=1}^4 \left( \vert A_j \vert ^2A_j\mathrm{e}^{-ik_{zj}z} \right) + \sum_{j=1, j \neq m}^4  2\, \vert A_j \vert ^2 A_m\mathrm{e}^{-ik_{zm}z}  + \sum_{j=1, j \neq m}^4 A_j^2A_m^* \mathrm{e}^{-i(2k_{zj}-k_{zm})z}  \label{p_NL_no_cc} \\ \nonumber & +  \sum_{j=1, j < l < m}^4 2\  \big(A_jA_lA_m^* \mathrm{e}^{-i(k_{zj}+k_{zl}-k_{zm})z} + A_jA_l^*A_m \mathrm{e}^{-i(k_{zj}-k_{zl}+k_{zm})z}+A_j^*A_lA_m \mathrm{e}^{-i(-k_{zj}+k_{zl}+k_{zm})z} \big) 
 \end{align} 
 \end{widetext}

 The first two terms of Eq. (\ref{p_NL_no_cc}) describe Self-Phase-Modulation (SPM) and Cross-Phase-Modulation (XPM). The other terms describe Four-Wave-Mixing (FWM), with the first term  $A_j^2A_m^*\mathrm{e}^{-i(2k_{zj}-k_{zm})z}$  being the degenerate process where the two pump waves are identical while the other terms being non-degenerate processes. 
 
 Our aim is to study the generation of new spectral components and their evolution along the propagation direction. In the following, we consider the waves $ \textbf{E}_1 $ and $ \textbf{E}_2 $ as the high-intensity pump waves whereas $ \textbf{E}_3 $ and $ \textbf{E}_4 $ correspond to the signal and idler waves respectively.  
 We describe the evolution of the spatial spectrum with propagation as a cascading process where $\textbf{E}_3$ is generated first by cross interaction of the pump waves. Then, four-wave mixing will amplify  $\textbf{E}_3$ and  $\textbf{E}_4$.  This approach is consistent with  the scaling performed by Gadonas {\it et al} \cite{Gadonas2001}.
 
 We separate the terms of the nonlinear polarization in Eq. (\ref{p_NL_no_cc}) according to the content of the exponents (which is equivalent to momentum conservation).  This leads to:
 
 \begin{widetext}
\begin{subequations}
\begin{align}
p_{NL,\, \omega_0}^{(1)}\mathrm{e}^{ik_{z1}z}= &  \label{P_NL1} \left( |A_1 |^2+ 2\, | A_2|^2  \right) A_1 \\
%%%%
p_{NL,\, \omega_0}^{(2)}\mathrm{e}^{ik_{z2}z}= &  \label{P_NL2} \left( |A_2 |^2+ 2\, | A_1|^2   \right) A_2 \\ 
%%%%
p_{NL,\, \omega_0}^{(3)}\mathrm{e}^{ik_{z3}z}= &  \label{P_NL3} 2\left(  | A_1|^2 +| A_2|^2  \right) A_3  +  A_1^2A_2^* \mathrm{e}^{-i(2k_{z1}-k_{z2}-k_{z3})z} + A_2^2A_1^* \mathrm{e}^{-i(2k_{z2}-k_{z1}-k_{z3})z} \\ \nonumber & +  A_1^2A_4^* \mathrm{e}^{-i(2k_{z1}-k_{z3}-k_{z4})z} + A_2^2A_4^* \mathrm{e}^{-i(2k_{z2}-k_{z3}-k_{z4})z} \\ \nonumber & + 2\,  \big(A_1A_2A_4^*\mathrm{e}^{-i(k_{z1}+k_{z2}-k_{z3}-k_{z4})z}  + A_1A_2^*A_4\mathrm{e}^{i(-k_{z1}+k_{z2}+k_{z3}-k_{z4})z}  + A_1^*A_2A_4\mathrm{e}^{i(k_{z1}-k_{z2}+k_{z3}-k_{z4})z} \big) \\
%%%%
p_{NL,\, \omega_0}^{(4)}\mathrm{e}^{ik_{z4}z}= &  \label{P_NL4} 2\left(  | A_1|^2 +| A_2|^2  \right) A_4  \\ \nonumber & +  A_1^2A_3^* \mathrm{e}^{-i(2k_{z1}-k_{z3}-k_{z4})z} + A_2^2A_3^* \mathrm{e}^{-i(2k_{z2}-k_{z3}-k_{z4})z} \\ \nonumber & + 2\,  \big(A_1A_2A_3^*\mathrm{e}^{-i(k_{z1}+k_{z2}-k_{z3}-k_{z4})z}  + A_1A_2^*A_3\mathrm{e}^{-i(k_{z1}-k_{z2}+k_{z3}-k_{z4})z} + A_1^*A_2A_3\mathrm{e}^{-i(-k_{z1}+k_{z2}+k_{z3}-k_{z4})z} \big)
\end{align} 
\end{subequations}
\end{widetext}
 
 For $p_{NL,\, \omega_0}^{(1)}$ and $p_{NL,\, \omega_0}^{(2)}$, we have neglected all contributions of $A_{3,4}$ as they are much smaller than $A_{1,2}$. Similarly, for $p_{NL,\, \omega_0}^{(3)}$ and $p_{NL,\, \omega_0}^{(4)}$, we have neglected all terms scaling with $A_{3,4}^2$.
 We note that the two contributions $A_1^2A_2^* \mathrm{e}^{-i(2k_{z1}-k_{z2}-k_{z3})z} $ and  $A_2^2A_1^* \mathrm{e}^{-i(2k_{z2}-k_{z1}-k_{z3})z}$ (first line of Eq. (\ref{P_NL3})) will be particularly important for the rest of   this paper. They arise from the XPM-like interaction with the pump, and scale as third power of the pump field and generate non-phase matched spectral broadening around the pump. The other 5 last terms scale with the second power of the pump and will contribute to amplification via four-wave mixing processes.  With this expression for the nonlinear polarization, we can now use the wave equation to derive the evolution of the fields.

% We have attributed/ associate two terms to $p_{NL,\, \omega_0}^{(3)}$ such as to describe the fact that $\textbf{E}_3$ is generated first by cross interaction with the pump. Those terms are scaling with third power of the pump $A_0$. These are: $A_1^2A_2^* \mathrm{e}^{-i(2k_{z1}-k_{z2}-k_{z3})z} $ and  $A_2^2A_1^* \mathrm{e}^{-i(2k_{z2}-k_{z1}-k_{z3})z}$ (first line of Eq. (\ref{P_NL3})). These two terms will be particularly important for the rest of this paper. 
 
 In the expression $p_{NL,\, \omega_0}^{(3)}$, the first two terms are cross-phase modulation terms that do not contribute efficiently to new spectral frequency generation in our case. The third and fourth, just mentioned above, are in third power of the pump while the rest of the nonlinear polarization terms are in second order of the pump. In the following, we will see that third and fourth terms will generate a crucial non-phase matched spectral broadening around the pump while the other terms will contribute to amplification via four-wave mixing processes.
 As the terms corresponding to the spectral broadening are more efficient, these will be the drivers for the cascade of four wave mixing.  Now, we can use the wave equation to derive the evolution of the fields.
 
\subsection{Evolution of the spatial spectrum along the propagation} 
The starting point is the scalar wave equation describing the full field $\sum_{j=1}^4 \textbf{E}_j$. We separate the full-field wave equation into four independent equations by using the separation approach described above:  
\begin{equation}
\Delta \textbf{E}_j- \dfrac{ \varepsilon _r}{c^2}\dfrac{ \partial ^2\textbf{E}_j}{ \partial t^2}= \mu _0\dfrac{ \partial ^2\textbf{P}^{(j)}_{NL}}{ \partial t^2}
\label{dEj}
\end{equation}

We now follow the approach in Ref. \cite{Tewari1996}: for each of the waves, we will only consider ideal Bessel beams defined by the $J_0$ Bessel function. To further simplify the analysis, we consider that the pump amplitude is undepleted with propagation. This, of course, implies that our analysis will only be valid for propagation distances shorter than the typical depletion scale length (in the examples shown, typically on the order of 1000~$\mu$m). The envelopes of the four interacting waves $A_j$ are then written as follows:
\begin{align}
A_j = a_j(z)\, J_0(k_{rj}r)
\end{align} 
where $k_{rj}$ is the transverse spatial frequency of the envelope $A_j$. %Since we are interested in the axial behavior of new spectral components $A_3$ and$A_4$, we considered the complex axial envelope parameter $a_3(z)$ and $a_4(z)$ defined in the same manner as in Ref. \cite{Tewari1996}. As for the pump waves, their axial envelopes will only vary in phase. 
Since we developed an expression of $\textbf{P}^{(j)}_{NL}$ with the same form as $\textbf{E}_j$, we can develop Eq. (\ref{dEj}) without the complex conjugate terms \cite{Tewari1996}. It becomes:
 \begin{equation}
2ik_{zj}\dfrac{\partial a_j(z)}{\partial z}J_0(k_{rj})=\dfrac{k_0^2}{\varepsilon _0}p^{(j)}_{NL,\omega _0}
\label{da4_z1}
\end{equation}
We multiply both parts of Eq. (\ref{da4_z1}) by $rJ_0(k_{r4}r)$ and integrate over 0 to $r_f$ which denotes the upper integration limit such as $r_f\gg 1/k_{r0}$. In other words, we perform Hankel transformation and thus study the evolution of the waves in Fourier space. The upper integration boundary $r_f$ is chosen finite so as to avoid infinite integrals or to avoid the introduction of apodization functions such as in Ref. \cite{Gadonas2001}. According to Ref. \cite{DLMF}: $$\int_0^{r_f}r J_0^2(k_{rj}r)\mathrm{d}r = r_f^2/2(J_0^2(k_{rj}r_f) + J_1^2(k_{rj}r_f) ) $$
 
 Then using the asymptotic expressions of both Bessel functions \cite{DLMF}, i.e. $$J_{\alpha}(k_{rj}r) = \sqrt{\dfrac{2}{\pi k_{rj}r}} \cos \left(k_{rj}r - \dfrac{\alpha \pi}{2} - \dfrac{\pi}{4} \right), \alpha =0,\, 1$$
 \noindent  This integral can be approximated to:
\begin{equation}
\int_0^{r_f}r J_0^2(k_{rj}r)\mathrm{d}r \approx \dfrac{r_f}{ \pi k_{rj}}
\label{Lommel2}
\end{equation}

\noindent which is proportional to $1/k_{rj}$, in agreement with Ref. \cite{Tewari1996}. 

Using the simplifications described in the previous section: $A_1 \approx A_2 \approx A_0=a_0 J_0(k_{r0}r)$ and $k_{z1}\approx k_{z2} \approx k_{z0}$, where $k_{z0} = k \cos{\theta}$ with $\theta$ the cone angle of the pump Bessel beam, our system of equations becomes:

\begin{widetext}
\begin{subequations}
\begin{align}
\dfrac{\partial a_0(z)}{\partial z}  = & -3i C_0  \tan(\theta_0) I_{TPM}^{(00)} a_0(z) \label{da0} \\ \dfrac{\partial a_3(z)}{\partial z} =& -iC_0 \tan(\theta_3) \Big(4 I_{TPM}^{(33)}  a_3(z) +2  I_{TPM}^{(03)} \mathrm{e}^{-i(\Delta k_{03}+\Phi_0^{NL})z} \label{da3} \\ \nonumber & \hphantom{{}-i C_0 \tan(\theta_3)  [} + 4 I_{TPM}^{(34)}  a_4^*(z)\mathrm{e}^{-i(\Delta k_{034}+2\Phi_0^{NL})z} +4I_{TPM}^{(34)} a_4(z)\mathrm{e}^{i\Delta k_{34}z} \Big) \\ \dfrac{\partial a_4(z)}{\partial z} =& -iC_0 \tan(\theta_4) \Big(4 I_{TPM}^{(44)}  a_4(z)  \label{da4} \\ \nonumber & \hphantom{{}-i C_0 \tan(\theta_4)  [} + 4 I_{TPM}^{(34)} a_3^*(z)\mathrm{e}^{-i(\Delta k_{034}+2\Phi_0^{NL})z} +4I_{TPM}^{(34)} a_3(z)\mathrm{e}^{-i\Delta k_{34}z} \Big)
\end{align}
\end{subequations}
\end{widetext}

\noindent where $C_0 = \pi \dfrac{k^2}{r_f} \dfrac{n_2}{n_0} I_0$. Here we write the solution of Eq. (\ref{da0}) as: $a_0(z) = \sqrt{I_0}\mathrm{exp}\left(-i\Phi_0^{NL} z\right)$ where $I_0$ is the peak input intensity. Solving Eq. (\ref{da0}) we obtain:

\begin{equation}
\Phi_0^{NL}=3 C_0  \tan(\theta_0) I_{TPM}^{(00)}
\end{equation}

\noindent The Transverse Phase Matching (TPM) integrals $I^{(j)}_{TPM}$ are defined as:

\begin{subequations}
\begin{align}
I_{TPM}^{(00)} & = \int_{0}^{r_f} J_0^4(k_{r0}\, r)\, r\, \mathrm{d}r \label{I_TPM_0} \\
I_{TPM}^{(03)} & = \int_{0}^{r_f} J_0^3(k_{r0}\, r)\, J_0(k_{r3}\, r)\, r\, \mathrm{d}r \label{I_TPM_1st} \\
I_{TPM}^{(34)} & = \int_{0}^{r_f} J_0^2(k_{r0}\, r)\, J_0(k_{r3}\, r)\, J_0(k_{r4}\, r)\, r\, \mathrm{d}r \label{I_TPM_2nd} \\
I_{TPM}^{(33)} & = \int_{0}^{r_f} J_0^2(k_{r0}\, r)\, J_0^2(k_{r3}\, r)\, r\, \mathrm{d}r \label{I_TPM_3} \\
I_{TPM}^{(44)} & = \int_{0}^{r_f} J_0^2(k_{r0}\, r)\, J_0^2(k_{r4}\, r)\, r\, \mathrm{d}r \label{I_TPM_4}
 \end{align} 
\end{subequations}

\noindent The indices relate to the waves involved in the last two Bessel functions in the integrals. Longitudinal wavevector mismatch terms are defined as: $\Delta k_{03}=k_{z0}-k_{z3}$, $\Delta k_{034}=2k_{z0}-k_{z3}-k_{z4}$ and $\Delta k_{34}=k_{z3}-k_{z4}$. The longitudinal phase matching conditions (when the wavevector mismatch defined above  equals zero) are the same as those reported in Ref. \cite{Gadonas2001}: the first one corresponds to FWM interaction of the proposed ``first approximation'' of the nonlinear Schr\"odinger Equation (NLSE). For this approximation, the same TPM integral as $I_{TPM}^{(03)}$ was also defined. Similarly, the other two Longitudinal Phase Matching (LPM) conditions were defined for the proposed ``second approximation'' of the NLSE along with $I_{TPM}^{(34)}$.

Hence, the signal and idler waves of our model can be assimilated to solutions of the first and second order approximations of the NLSE of Ref. \cite{Gadonas2001}. Our target here is to build a fully explicit model where the mechanism actually driving the generation of new spectral components can be analytically identified. However, before integrating Eqs. (\ref{da3}) and (\ref{da4}), we first show numerical results so that we can later compare analytical results with numerical modelling results of the full NLSE.

\section{Dynamics of the growth of new spectral components}

\subsection{Numerical model and results}

Our numerical simulations are based on the nonlinear Schr\"odinger equation (NLSE) given in Refs. \cite{Gadonas2001, Couairon2011} for a monochromatic beam propagating in a Kerr medium \label{NLSEModel}:  

\begin{align}
\dfrac{\partial A}{\partial z}=\dfrac{i}{2k}\Delta_{\bot}A + \dfrac{ik\, n_2}{n_0}|A|^2A \label{model1}
\end{align}

\noindent where $A$ is the linearly polarized complex amplitude of the laser electric field, $\Delta_{\bot}=1/r\partial/\partial r + \partial^2 /\partial r^2$ is the transverse Laplacian operator, $r$ and $z$ are the radial and axial coordinates, $k$ is the wavevector in the medium, $n_0$ and $n_2$ are the linear and nonlinear refractive indices. Parameters of our simulations are given in Table \ref{tab1} and correspond to the realistic propagation of a high-intensity pulse in fused silica. Since nonlinear instabilities stem mainly from Kerr nonlinearities, we neglected other nonlinear effects (particularly nonlinear losses which are known attenuate nonlinear instabilities \cite{Porras2004,Polesana2008}) so as to isolate the effect of intensity shaping on the control of nonlinear instabilities in Bessel beams.

\begin{table}[b]
\caption{\label{tab1}%
Numerical parameters used in simulations}
\begin{ruledtabular}
\begin{tabular}{cc}
$\lambda$ ($\mu$m) & $0.8$\\
$n$ & $1.45$\\
$n_2$ (m$^2$/W) & $2.48\ 10^{-20}$\\
$\theta$ ($^{\circ}$) & $4$\\
$w_0$ ($\mu$m) & $300$\\
\end{tabular}
\end{ruledtabular}
\end{table}

The input field (a Bessel-Gauss (BG) beam) is modeled by a Gaussian beam with a spatial phase characterizing the axicon conical focusing: $A_{BG}(r,z=0)=A_0\mathrm{exp}\left( -r^2/w_0^2-k\, r\, \sin(\theta)\right)$ where $w_0$ is the input Gaussian beam waist \cite{Jarutis2000, Polesana2008}. 

\begin{figure}[htp]
\centering
\includegraphics[width=0.45\textwidth]{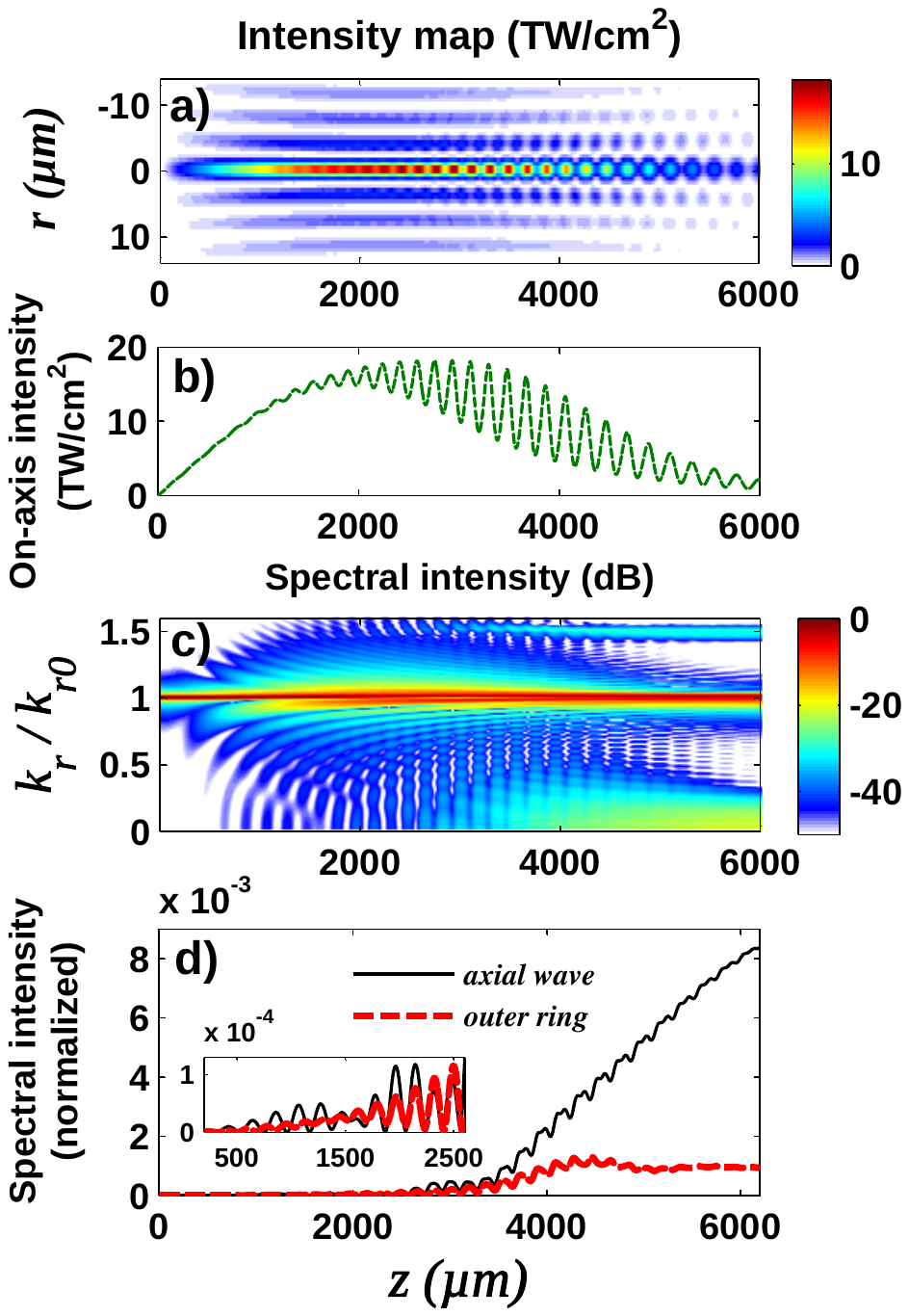}
\caption{\label{instability_BG} NLSE simulation results: (a) intensity distribution of a Bessel–Gauss (BG) beam propagating in a pure nonlinear Kerr medium as a function of the radial and propagation distances $r$ and $z$ (Input peak intensity of the Gaussian beam 33.7~GW cm$^{-2}$), (b) the corresponding on-axis intensity, (c) the spatial spectrum distribution $|\tilde{A}(k_{r})|^2/|\tilde{A}(k_{r0}|^2|$ (logarithmic scale dB) and (d) the spectral intensity of the axial wave (solid line) and outer ring in linear scale (dashed line) along the propagation distance. The intensity of both spectral components are normalized to the maximal intensity value of the central frequency.}
\end{figure}

 The results of the integration of Eq. (\ref{model1}), based on a split-step algorithm are shown in Fig. \ref{instability_BG}. In Fig. \ref{instability_BG}(a), we plot the evolution of the intensity as a function of  radial distance $r$ and propagation distance $z$. We observe that the beam intensity undergoes longitudinal modulation not only along the central core but also in the peripheral rings. The on-axis intensity, {\it i.e.} the intensity $I(r=0,z)$, is shown in Fig. \ref{instability_BG}(b), where the intensity oscillations, with a period of $\sim$$180.5~$$\mu$m, are clearly apparent.
 
  In Fig. \ref{instability_BG}(c), the spatial spectrum $|\tilde{A}(k_r,z)|^2$ is plotted as a function of the propagation distance. We recall that the spatial spectrum of a Bessel beam, in the linear regime, is in the form of a ring centered around the Bessel transverse frequency $k_{r0} = 0.8\ \mu$m$^{-1}$ \cite{Jarutis2000}. We display the evolution of a cross-section along the propagation distance. After an initial stage of spectral broadening around the central frequency $k_{r0}$, we notice the generation of two particular spectral components at $k_r \approx 0$ and $k_r \approx 1.5 k_{r0}$. These components are respectively referred to as the axial wave and outer ring ($k_r\sim\sqrt{2} k_{r0}$) as mentioned in the introduction and reported in previous works \cite{Gadonas2001,Polesana2007,Ouadghiri2017}. It is the interference of the input Bessel beam with the two new spectral components which generates the oscillations observed on the on-axis intensity distribution. (The interference pattern was initially interpreted only as interference between the Bessel beam and the axial wave, but we note that the secondary wave also generates interference with precisely the same period). 
  
In more detail, Fig. \ref{instability_BG}(c) shows two regimes. From a propagation distance range from 0 to $z \sim 2600$ $\mu$m, the spatial spectrum progressively broadens around the central frequency. It is only for further propagation distances that the growth of the axial wave and outer ring is efficient. We specifically show the evolution of these spectral components in Fig. \ref{instability_BG}(d). In Fig. \ref{instability_BG}(c), we also note parabolic-like structures for spatial frequencies around $k_{r0}$. Those were not discussed in previous literature and our analytical model will allow us to explain them.

\subsection{Analysis using the Four Wave Mixing model}

Here we show that the main characteristics of the first and second regime can be qualitatively described using the Four-Wave Mixing model developed in the previous section (Eqs. (\ref{da3},\ref{da4})). 

For very short propagation distances, since the amplitude of $a_4$ is near 0, as discussed in section \ref{sec:ThirdOrderPolar} above, we first neglect the terms in $a_4$ in the expression of the evolution of $a_3$ (Eq. (\ref{da3})).
 
 This becomes: 
 \begin{align}
   \dfrac{\partial a_3(z)}{\partial z} = -i C_0 \tan(\theta_3) & \Big[ 4 I_{TPM}^{(33)}  a_3(z) \label{da3_NOa4}   \\ \nonumber & +2 \sqrt{I_0} I_{TPM}^{(03)} \mathrm{e}^{-i(k_{z0}-k_{z3} + \Phi_0^{NL})z} \Big]
 \end{align}

 %\noindent We introduce $b_3 = a_3 e^{i \Phi_3^{NL} z}$ where $\Phi_3^{NL}=4 C_0 \tan(\theta_3) I_{TPM}^{(33)}$ such as:

%\begin{align}
%\dfrac{\partial b_3(z)}{\partial z} =& -2i C_0 \sqrt{I_0} \tan(\theta_3)  I_{TPM}^{(03)} \mathrm{e}^{-i(k_{z0}-k_{z3} + \Phi_0^{NL} -\Phi_3^{NL})z} \\ \nonumber
% \label{db3_NOa4}
%\end{align}

 Using $\Phi_3^{NL}=4 C_0 \tan(\theta_3) I_{TPM}^{(33)}$ and $\Delta k_{03}^{eff} = \Delta k_{03} + \Phi_0^{NL} -\Phi_3^{NL}$,  $a_3$ is given by:

\begin{align}
 a_3(z) = - & 2 i C_0 \sqrt{I_0} \tan(\theta_3)  I_{TPM}^{(03)} \mathrm{e}^{-i\left(\Delta k_{03}^{eff} z/2+\Phi_3^{NL} z\right)} \nonumber \\
 &  \times z\ \mathrm{sinc}\left(\Delta k_{03}^{eff} z/2\right)  \label{a3_NOa4_2} 
\end{align}

This result will be very important in the following sections of the paper. Then we can compute the evolution of the idler wave with Eq. (\ref{da4}), using the expression of the signal wave $a_3$ computed above.

Similarly, using, $\Phi_4^{NL}=4 C_0 \tan(\theta_4) I_{TPM}^{(44)}$ and $\Delta k_{04}^{eff} = k_{z0}-k_{z4} + \Phi_0^{NL} -\Phi_4^{NL}$,  $\Delta k_{034}^{eff}= \Delta k_{034} + 2\Phi_0^{NL} -\Phi_3^{NL} -\Phi_4^{NL}$, and $\Delta k_{34}^{eff}= \Delta k_{34} + \Phi_3^{NL} -\Phi_4^{NL}$. $a_4$ is given by:

\begin{align}
 a_4(z) = & -8i C_0^2 \sqrt{I_0} I_{TPM}^{(03)}I_{TPM}^{(34)} \dfrac{\tan(\theta_3)\tan(\theta_4)}{\Delta k_{03}^{eff}} \mathrm{e}^{-i\Phi_4^{NL} z} \nonumber  \\ \nonumber 
& \times \Big\{ 2\, \mathrm{exp}{\left(-i\Delta k_{04}^{eff} z/2\right)} z\ \mathrm{sinc}\left(\Delta k_{04}^{eff} z/2\right) \\ \nonumber &  - \mathrm{exp}{\left(-i\Delta k_{034}^{eff} z/2\right)} z\ \mathrm{sinc}\left(\Delta k_{034}^{eff} z/2\right) \\  &  - \mathrm{exp}{\left(-i\Delta k_{34}^{eff} z/2\right)} z\ \mathrm{sinc}\left(\Delta k_{34}^{eff} z/2\right) \label{a4_a3} \Big\} 
\end{align}
%\hphantom{{}-4i C_0 I_{TPM}^{2nd} \tan(\theta_4)  [}  + b_3(z)\mathrm{e}^{-i(k_{z3}-k_{z4}+4 C_0 \tan(\theta_3) I_{TPM}^{(3)})z} \big) 

\begin{figure}[hbtp]
\centering
\includegraphics[width=0.45\textwidth]{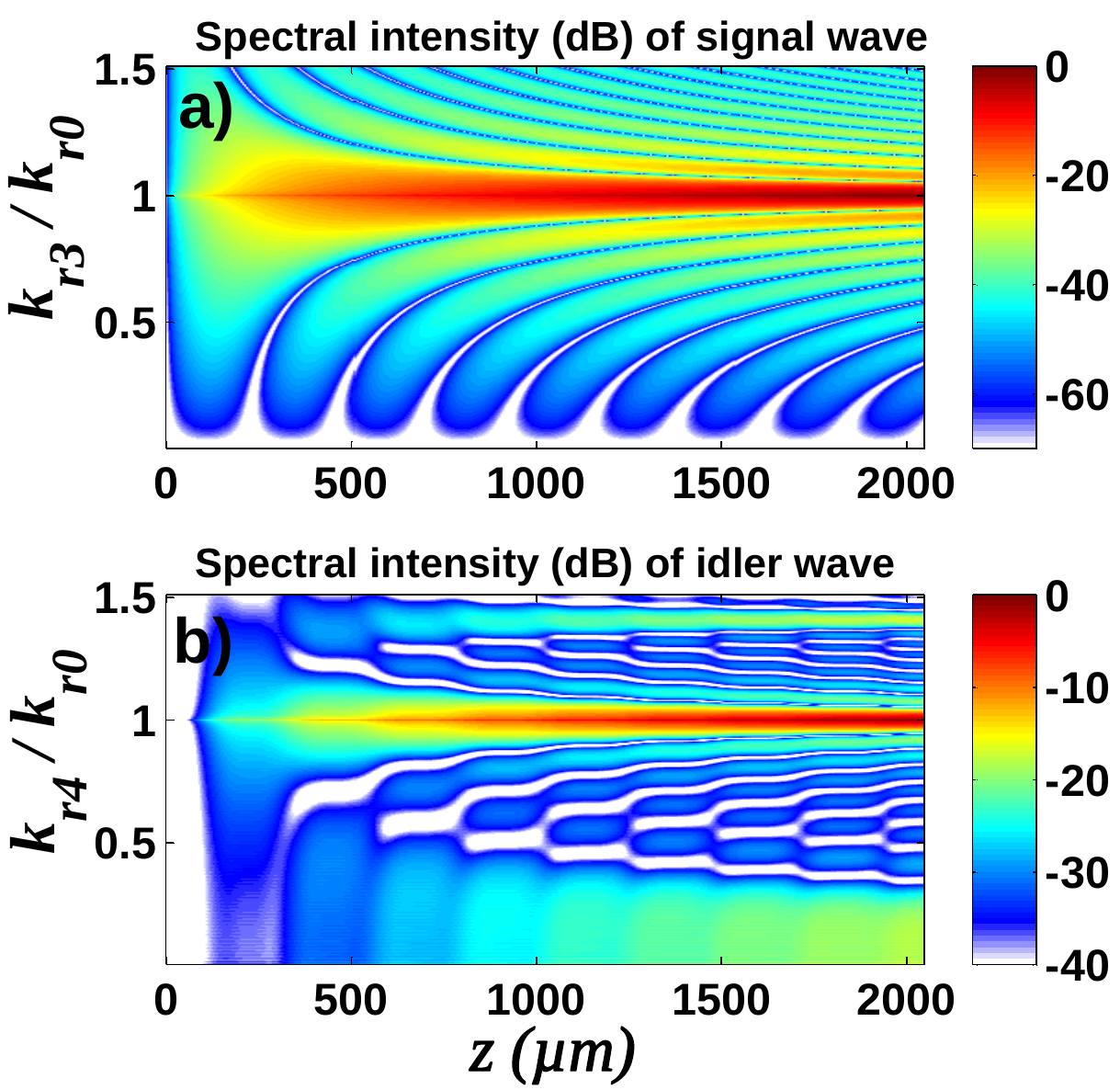}
\caption{\label{approx1} Evolution of frequency-resolved intensity of the (top) signal and (bottom) idler waves along propagation. The latter is computed for $\theta_3 = 0.005\theta_0$.}
\end{figure}

\subsubsection{First regime: Spatio-Spectral broadening characteristics}

Using Eq. (\ref{a3_NOa4_2}), we show in Fig. \ref{approx1}(a) the evolution of the intensity of $a_3(z)$ along the propagation direction for different values of $k_{r3}$. In the figure, the nonlinear phases $\Phi^{NL}_0$ and $\Phi^{NL}_3$ were evaluated for an intensity of $18$~TW/cm$^{2}$ as for the peak power in Fig. \ref{instability_BG}. We also choose the same pump cone angle $\theta=4^{\circ}$ as in our numerical simulation of the NLSE. 

 This regime qualitatively reproduces the first growth stage discussed earlier (propagation from $0$ to $z=2600$ $\mu$m).
The signal wave ($a_3$) oscillates for frequencies that are away from the central frequency $k_{r0}$ and exhibits parabolic-like fringe structure in the $k_r-z$ space, consistent with NLSE simulation results (Fig. \ref{instability_BG}(c)). These parabolic structures oscillate with a frequency-dependent period: $p(k_{r3})=2\pi/\Delta k_{03}^{eff}$ where we can neglect the nonlinear phase to obtain:

 $$p(k_{r3})\simeq \frac{2\pi}{|k_{z0}-k_{z3}|}\simeq \frac{4 \pi k}{|k_{r0}^2-k_{r3}^2|}$$

 This result is in very good quantitative agreement with the oscillation period observed in Fig. \ref{instability_BG}. For instance, for $k_{r3} = 0.5 k_{r0}$, the expression above gives a period of $302 ~\mu$m, while the numerical result is $295 ~\mu$m. 
 
\subsubsection{Second regime: Growth of new spectral components and interference pattern}
 
After this initial spectral broadening, numerical results show that both the axial wave and outer ring will be amplified for $z>2600~\mu$m. 
Now, we show that in this second regime, an axial wave ($k{r}\simeq 0$) and an outer ring ($k{r}\simeq \sqrt(2) k_{r0}$) are amplified. This amplification is described by the expression of $a_4$.

The transverse phase matching integral $I_{TPM}^{(34)}$ peaks for $k_{r3}\simeq 0$. Therefore, we plot in Fig.  \ref{approx1}(b) the frequency-resolved evolution of $|a_4(z)|^2$ for a signal wave $a_3$ at $k_{r3}$ close to zero. We observe the amplification of an axial wave ($k_{r4}\simeq 0$) and of an outer ring $k_{r4}\simeq \sqrt(2) k_{r0}$. The axial wave arises from the last term in Eq. (\ref{a4_a3}), for which $\Delta k_{34}^{eff} =0$ and the outer ring arises from the second to last term in the same equation, for which $\Delta k_{034}^{eff} =0$. 

We note that our description does not yet take into account the complete set of signal waves $a_3$ that are continuously generated in the first spectral broadening stage as the analysis would be extremely laborious.  

In summary, we have shown that our model can explain detailed features of the nonlinear propagation of Bessel-Gauss beams. We have seen that it is the initial broadening stage (generating the wave $a_3$) that determines the efficiency of the FWM-induced amplification in the subsequent stage. In the next section, we will expand our theory to take into account the spectral phase of the pump. To simplify our analysis, we will restrict ourselves to the first broadening regime.

\section{Reduced model}

Here we will describe the generation of $a_3$ when taking into account the fact that the input pump Bessel beam is spectrally extended. We will see in the next sections how the spectral phase will impact the growth of $a_3$.

We restart our analysis from Eq. (\ref{da3_NOa4}). In this expression, the first nonlinear term, $-4iC_0I_{TPM}^{(33)}\tan(\theta_3)a_3$, corresponds to cross-phase modulation and is much weaker than the second term. Therefore, to simplify our analysis, we neglect the first term in our reduced model. We also drop out the nonlinear phase terms.
Now, we take into account a pump beam described by: $A_0 = \sqrt{I_0}\int dk_r S(k_{r}) J_0(k_r r) e^{i\phi(k_r)}$ where $S(k_{rj})$ stand for the amplitude distribution of the spectral components of the pump and signal waves. The complex spatial spectra are given by: $ \tilde{S}(k_{r}) = S_j(k_{r})\mathrm{exp}[i\phi_j(k_{r})]$. This way, it is possible to take into account the input spectral distribution by associating with each spectral component the corresponding amplitude and phase values.  Equation (\ref{da3_NOa4}) becomes:
% \begin{widetext}
% \begin{align}
% \dfrac{\partial a_3(z)}{\partial z} = -2iC_0 \sqrt{I_0} \tan(\theta_3) \mathrm{e}^{ik_{z3}z}\int_{0}^{r_f} \nonumber J_0(k_{rj}\, r)\,  r\, \mathrm{d}r \Bigg\{ \left[\int dk_r S(k_{r}) J_0(k_r r) e^{i\phi(k_r)-ik_{z}z} \right]^2 \\  \left[\int dk_r S(k_{r}) J_0(k_r r) e^{-i\phi(k_r)+ik_{z}z} \right] \Bigg\} \label{da3_reduced} 
% \end{align}
% \end{widetext}

\begin{align}
\dfrac{\partial a_3(z)}{\partial z} = -2& iC_0 \sqrt{I_0} \tan(\theta_3) \mathrm{e}^{ik_{z3}z} \int_{0}^{r_f} J_0(k_{rj}\, r)\,   \nonumber  \\ \label{da3_reduced0} \times &  \Bigg\{ \left[\int dk_r S(k_{r}) J_0(k_r r) e^{i\phi(k_r)-ik_{z}z}\right]^2 \\ & \; \left[\int dk_r S(k_{r}) J_0(k_r r) e^{-i\phi(k_r)+ik_{z}z} \right]  \nonumber \Bigg\}  r\, \mathrm{d}r
\end{align}

Here, we have a triple integral over the transverse spatial frequency $k_r$ where $k_z=\sqrt{k^2-k_r^2}$. To make this expression easier to analyze analytically, we define for each of these integrals a different parameter, i.e. $k_{rj}$, $k_{rl}$ and $k_{rm}$. Our equation can then be written as:

\begin{align}
\dfrac{\partial a_3(z)}{\partial z}= &-2i C_0 \sqrt{I_0}\tan (\theta_3) \int\limits_{0}^{k}\mathrm{d}k_{rj}S(k_{rj}) \int\limits_{0}^{k}\mathrm{d}k_{rl}S(k_{rl}) \label{da3_reduced} \\ \nonumber & \times \int\limits_{0}^{k}\mathrm{d}k_{rm}S(k_{rm})\, I_{TPM}^{(jlm)}\, \mathrm{exp}{(i\Delta \Phi_{jlm} - i\Delta k_{jlm}z)}  
\end{align}

\noindent where $I_{TPM}^{(jlm)}= \int_{0}^{r_f} J_0(k_{rj} r) J_0(k_{rl} r)J_0(k_{rm} r) J_0(k_{r3} r) r\mathrm{d}r$, $\Delta \Phi_{jlm}=\phi(k_{rj}) + \phi(k_{rl})- \phi(k_{rm})$ and $\Delta k_{jlm}= k_{zj} + k_{zl} - k_{zm} - k_{z3}$.

Now, this model will allow us to predict how the efficiency of the first spectral broadening stage is affected by the spectral phase of the input pump beam. In the next two sections, we will use our model to understand two different cases of the literature regarding the propagation of Bessel beams in Kerr media.
%%%%%%%%%%%%%%%%%%%%%%%%%%%%%%%%%%%%%%%%%%%%%%%%%%%%%%%%%%%%%%%%%%%%%%%%%%%%%%%%%%%%%%%%%%%%%%%%%%%%%%%%%%%%%%%%%%%%%%
\section{Soft or abrupt input conditions}
Previous work by other groups \cite{Sogomonian1999,Polesana2007,Dubietis2007} showed experimentally and numerically that an abrupt transition between linear and nonlinear propagation of an intense Bessel beam yields efficient generation of outer ring and axial wave components. In contrast, this is much less efficient when the Bessel beam is smoothly forming into the nonlinear medium (see Fig. 1 in Ref. \cite{Dubietis2007}). These two input conditions are respectively referred to as soft and abrupt input conditions. %When nonlinear losses are included, this can bee seen as the transition between steady and non-steady filamentation.  An explanation was provided by the stability of the NonLinear Unbalanced Bessel Beams (NLUBB) \cite{Polesana2008}, but this requires nonlinear losses, which are not considered here. 

\begin{figure*}[htb]
\centering
\includegraphics[width=\textwidth]{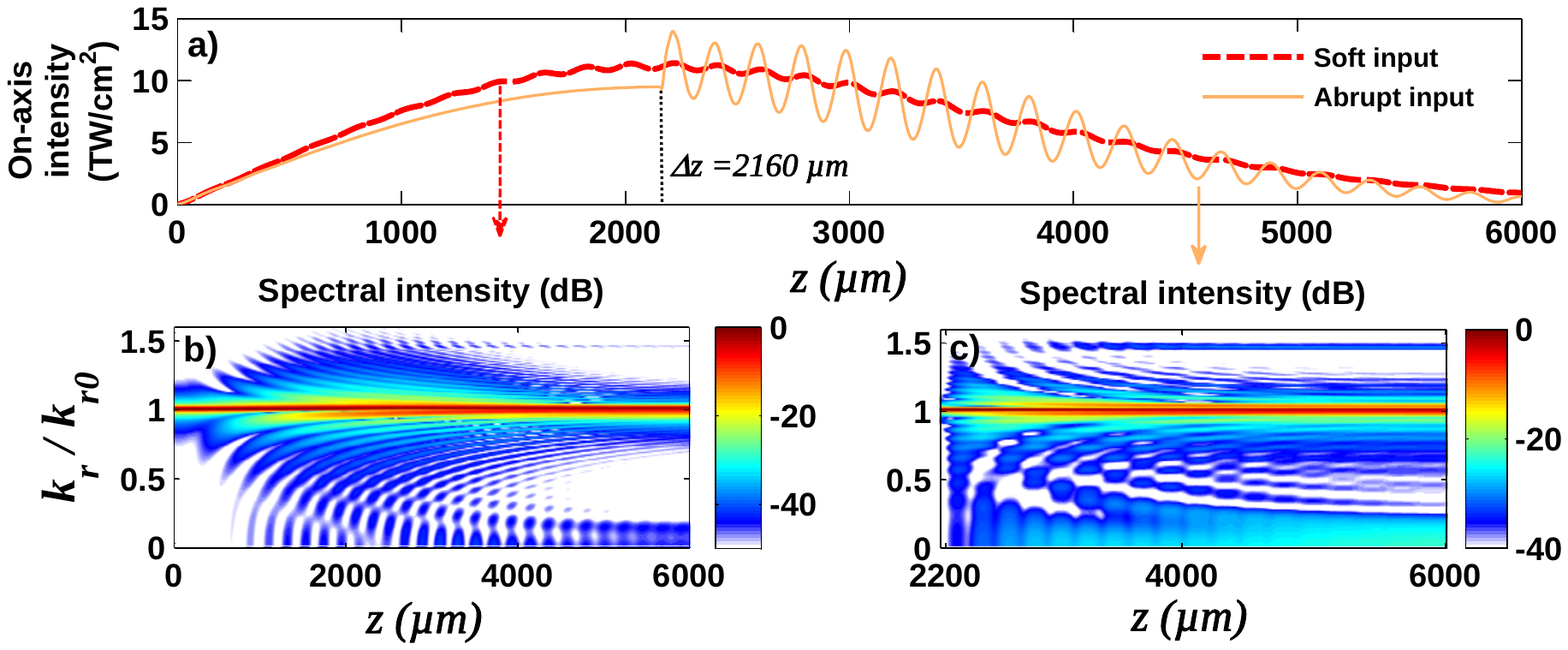}
\caption{\label{input_sim} (a) On-axis intensity distribution of a Bessel–Gauss beam propagating in a pure nonlinear Kerr medium for the soft (dashed line) and abrupt (dotted line) inputs; (b-c) the corresponding spatial spectra evolution along the propagation distance (logarithmic scale) for both cases respectively. Note that the intensity of newly-generated spectral components in the case of abrupt transition is two order of magnitude higher than the one reached in case of soft transition. Parameters are the presented in table \ref{tab1} and $P_{in}=31.2$ MW.}
\end{figure*}

In Fig. \ref{input_sim}, we show numerical results of the NLSE, that includes only Kerr effect, as described in section \ref{NLSEModel}.

We use the same parameters as in table \ref{tab1} except for the input power which was reduced to $P_{in}=31.2$ MW, corresponding to a beam peak intensity of $I_{max} = 9$~TW/cm$^{2}$. 
In the linear regime, this Bessel beam reaches its peak intensity at $z = 2160$ $\mu$m. For the abrupt input condition, the nonlinear medium starts at this point, whereas for the soft input conditions, the nonlinear propagation starts at $z=0$. 

We compare the evolution of on-axis intensity and spatial spectrum for soft and abrupt input conditions for the same Bessel beam. We see, in agreement with literature, that the abrupt transition generates pronounced on-axis intensity modulation, in stark contrast with soft input conditions. The evolution of spatial spectra can be compared from Figs. \ref{input_sim}(b-c). For the abrupt input condition (c), the spectral intensity of the axial wave and outer ring components quickly grows, with an intensity two orders of magnitude higher than in the case of the soft input condition (b). We note that the case of soft input condition physically corresponds to the same case as previous sections. The oscillations of the on-axis intensity are reduced because of a smaller input power.

\begin{figure}[htb]
\centering
\includegraphics[width=0.45\textwidth]{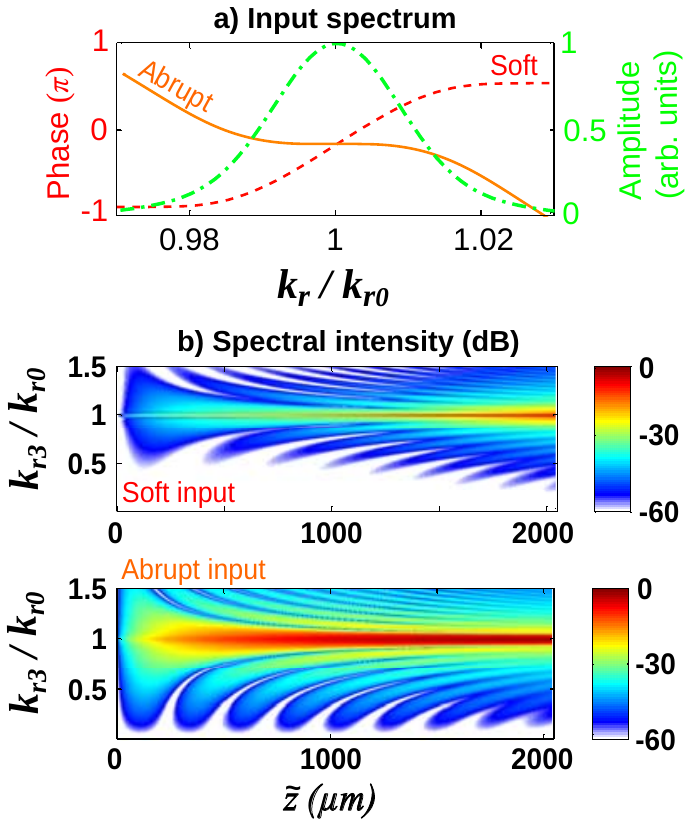}
\caption{\label{BG_diff_z} a) Comparison between the input phase distributions at $\tilde{z}= z-\Delta z=0$) where $\Delta z$ marks the position of the beam with sample input, where it is 0 for the soft case (red dashed line) and 2160~$\mu$m for the other case of abrupt input condition (red dotted line)  b) Results using our FWM model: (spatial) frequency- resolved intensity (dB) of the signal wave resulting from the interaction of three pump waves composed each of five frequencies: $k_r/k_{r0} = [ 0.96\ -\ 1.04]$ in case their respective spectral phases correspond to the (top) soft and (bottom) abrupt input conditions respectively as described in Fig. \ref{BG_diff_z}(a).}
\end{figure} 

We will now use our reduced model to understand the different behaviors. We demonstrate that the spectral phase distribution largely differs between soft and abrupt input conditions, which impacts on the first stage of spectral broadening.

% In the soft transition case, even though the peak intensity is about the same as the one reached in the abrupt case at $z=2160$ $\mu m$, the axial wave only present an oscillating behavior and no noticeable growth is observed. Considering that the intensity of the axial wave seed is higher by two order of magnitudes in the abrupt transition case, we assume that it leads to more efficient FWM interactions. We conclude that the main point of difference between the soft and abrupt transition cases lies in the initial spectral broadening induced by SPM. While the latter is being gradually established in the soft transition case, it encompasses nearly all frequencies in the other case. \

  We compare in Fig. \ref{BG_diff_z}(a) the input spectral amplitude (green) and phase (red) corresponding to both cases. The distance $\tilde{z}$ is the relative distance to the transition point between the linear and nonlinear medium. While the input amplitude is naturally the same (in the form of a Gaussian), we notice that the phase distribution is much steeper in the soft input condition case. In contrast, it is quasi-flat within the spectral range around the peak of the amplitude. %The input amplitude takes the form of a Gaussian function and is mostly in the range $\k_r/k_{r0} \in [0.98\, -\, 1.02]$). \ %In the following, the Bessel ring will be defined in this spectral range. \
  
A quasi-flat spectral phase implies that spectral components composing the pump wave are nearly in-phase. Qualitatively, if each of these spectral components interacts according to the four-wave mixing process described above, then each signal wave generated from these interactions will be in phase with the others. The resulting signal wave at a given frequency will then be made of constructive interference between all these waves, which explains the very rapid growth of spectral components at about all frequencies around the central one in the abrupt input condition.  In contrast, a steep spectral phase profile, which implies out-of-phase spectral components, leads to partially-destructive interferences and the axial wave  will then be weaker. 
Now, we use Eq. (\ref{da3_reduced}) to obtain an analytical explanation. The triple integral over the spectrum is unfortunately too heavy and for sake of simplicity, we restrict the pump beam to only two spectral components defined at $k_{ra}$ and $k_{rb}$, such that $k_{ra} \approx k_{rb} \approx k_{r0}$, with $\phi_a$ and $\phi_b$ being their respective input spectral phases and consider they have the same amplitude. The spectral distribution of each of interacting waves can then be written as $ \tilde{S} (k_{r})= \delta(k_{r} -k_{r0}) \, [\mathrm{exp}(i \phi_a) + \mathrm{exp}(i \phi_b)]$. 

The signal wave intensity $I_3(z)=|a_3(z)|^2$ is then found to be proportional to:
\begin{align}
I_3(z) \propto \ & \left(\tan(\theta_3)\, I_{TPM}^{(03)} \right)^2 z^2\,  \mathrm{sinc}^2[\Delta k_{03}z/2] \label{I_3_SPM} \\
& \times [1+ \cos(\phi_{a}-\phi_{b})]^3. \nonumber
\end{align}

The growth of the signal wave intensity is proportional to the term $z^2\mathrm{sinc}^2[\Delta k_{03}z/2]$ which indicates the above discussed oscillating behavior. Of particular interest, $I_3(z,k_{z3})$ is proportional to the cube of the phase-dependent term $[1+ \cos(\phi_{a}-\phi_{b})]$ which shows that non-zero phase difference quickly reduces the peak value of the oscillations and thus decreases the magnitude of the axial wave seed. \

In Fig. \ref{BG_diff_z}(b), we numerically solve our reduced FWM equation (\ref{I_3_SPM}) over a more realistic case of a pump spectrum composed of five frequencies in the range $k_r/k_{r0} \in [ 0.96\ -\ 1.04]$. We plot the evolution of the signal wave spectral intensity along propagation and compare results in soft and abrupt input conditions. For each case, the input spectral phase distribution is extracted from Fig. \ref{BG_diff_z}. 
In the soft input condition, initial spectral broadening is very weak and the generated frequencies are close to the central one. Notice that it gradually extends to more frequencies for longer propagation distances in cascaded-like fashion, in good qualitative agreement with numerical simulations of Fig. \ref{input_sim}(b) in the propagation range $[0 \ - \ 1500]~\mu$m. In contrast, in the case of abrupt input conditions, the spectrum very rapidly broadens, again in agreement with Fig. \ref{input_sim}(c). 

Therefore, we conclude that the spectral phase is an effective control parameter for the initial broadening regime and therefore a control parameter for the instabilities occurring in the second stage.

%%%%%%%%%%%%%%%%%%%%%%%%%%%%%%%%%%%%%%%%%%%%%%%%%%%%%%%%%%%%%%%%%%%%%%%%%%%%%%%%%%%%%%%%%%%%%%%%%%%%%%%%%%%%%%%%%%%%

\section{Control of nonlinear instabilities using shaped intensity profiles}

\begin{figure*}[ht]
\centering
\includegraphics[width=0.95\textwidth]{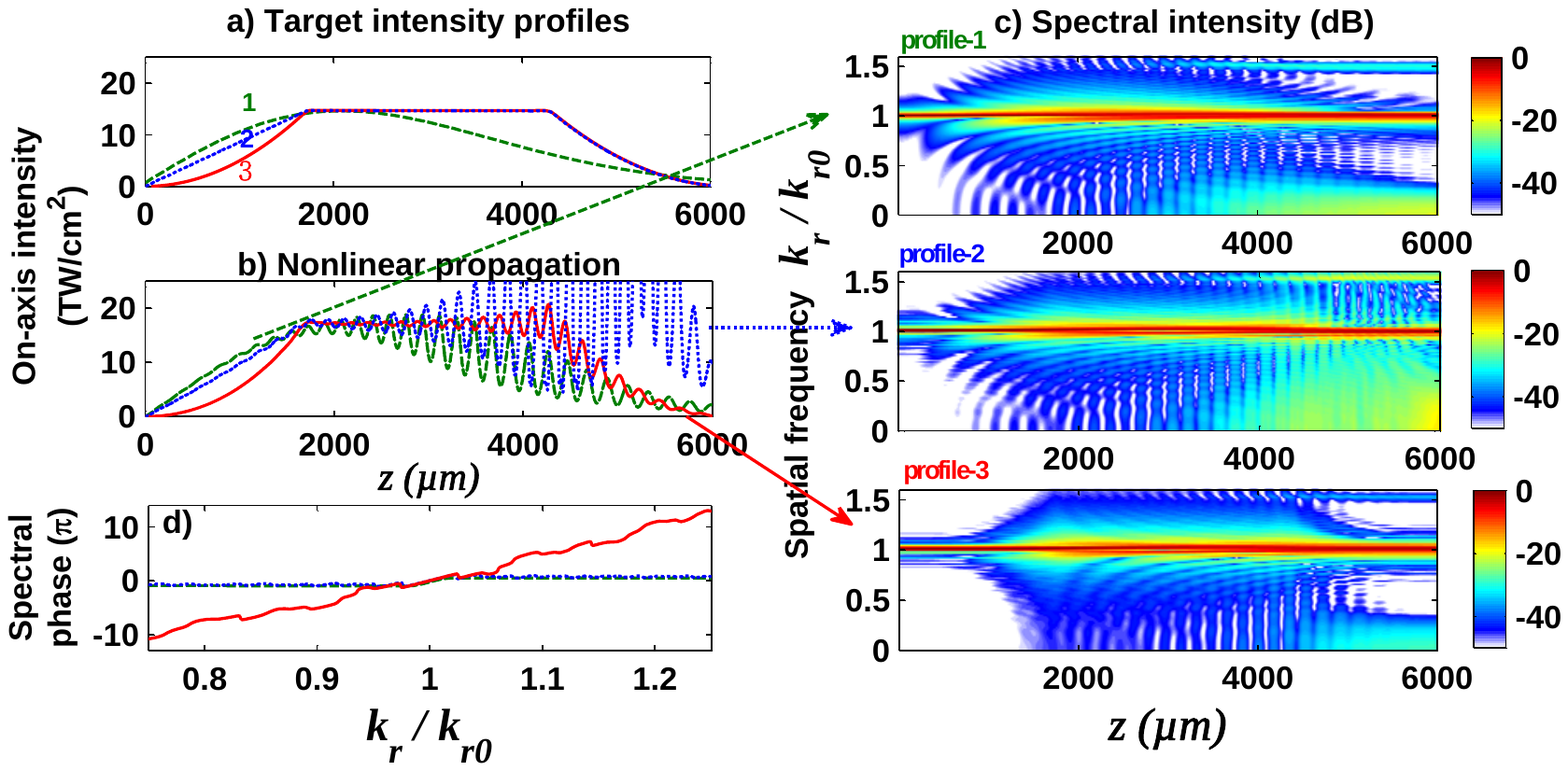}
\caption{\label{BB_shape_sim} Simulation of the nonlinear propagation of three Bessel beams with different (a) target on-axis intensity profiles. Evolution of their respective (b) on-axis intensities, (c) spatial spectra (dB) along propagation. Spectral intensities are normalized to their respective maximal values. Numerical parameters are described in table \ref{tab1}. (d) Input spectral phase profiles corresponding to the three Bessel beams.}
\end{figure*}

In this section, we will interpret previous numerical results on the control of nonlinear instabilities depending on the initial intensity rise in Bessel beams propagating inside nonlinear Kerr media. In Ref. \cite{Ouadghiri2017}, we compared the nonlinear propagation of three Bessel beams with different on-axis  intensity profiles in the linear propagation regime.  Their peak maximal intensity is chosen to be the same in order to study the effect of the initial intensity rise on the growth of nonlinear instabilities. These three target intensity profiles are depicted in Fig. \ref{BB_shape_sim}(a) and are described as follows: the first profile, denoted profile 1, is that of a conventional Bessel-Gauss beam (green dashed line) identical to that we have used previously. The second profile (profile 2) consists of a linear leading edge followed by a flat-top intensity and parabolic decaying trailing edge (blue dotted line). Profile 3 is identical to profile 2 except that it exhibits a parabolic intensity rise instead of a linear ramp (red solid line). Numerical parameters are the same as in table \ref{tab1}. \

 We used the same nonlinear propagation equation described by Eq. (\ref{model1}). The evolution of the on-axis intensity  of these Bessel beams is presented in Fig. \ref{BB_shape_sim}(b). Compared to the case of the BG beam (profile 1), the two other beams present oscillations of the central core intensity that has the same period, but their amplitudes strongly differ. Although profiles 2 and 3 only differ in the initial intensity rise, the oscillation amplitude is significantly smaller for profile 3. 
 
In the spectral domain (Fig. \ref{BB_shape_sim}(c)), the observed weak on-axis intensity oscillations in case of profile 3 correspond to weak intensity growth of the axial wave ,  below -40~dB up to $z=$ 4000 $\mu$m and its amplification remains around -30 $\mathrm{dB}$ afterwards. In addition, in the initial stage of spectral broadening, in the range $z=[0-2000]$~$\mu$m, we observe strong oscillations in the spectrum for profiles 1 and 2 while this oscillating behavior is initially absent in the case of profile 3 and only appears at a propagation distance around $ z=3000 $ $\mu$m. Following the appearance of these oscillations, both the axial wave and outer ring increase in intensity which indicates that FWM processes become active past this propagation point. However, their growth remains noticeably weaker compared to the other two Bessel beams.  \

%\begin{figure*}[htb]
%\centering
%\includegraphics[width=\textwidth]{fig_chap4/BB_shape3.pdf}
%\caption{\label{BB_shape_sim} Simulation of the nonlinear propagation of three Bessel beams with different (a) target on-axis intensity profiles. Evolution of their respective (b) on-axis intensities, (c) spatial spectra (dB) and (d) spectral intensities of axial wave (linear scale) along propagation. Spectral intensities are normalized to their respective maximal values. Numerical parameters are described in table \ref{tab31}. }
%\end{figure*}

In Ref. \cite{Ouadghiri2017}, we reported that it is not because profile 3 has initially a low-intensity zone that nonlinear instabilities are weaker. It is, as it is the case with soft and abrupt input conditions, the input spectral phase that is main factor of influence. Figure (\ref{BB_shape_sim},d) shows the spectral phase distributions corresponding to profiles 1, 2, and 3. The BG beam (profile 1) has a spectral phase  in the form a linear ramp in the range ($k_r/k_{r0} \in [0.98\, -\, 1.02]$) and is flat outside this range. The spectral phase distribution of profile 2 is similar to that of profile 1 except that it exhibits very weak oscillations  in the tails of the spectrum. In the case of profile 3, however, while the spectral phase also takes the form of a linear ramp in the range $k_r/k_{r0} \in [0.98\ -\ 1.02]$, it varies significantly outside, with a quasi-linear ramp. \

 The spectral domain of influence of the phase is the range $k_r/k_{r0} \in [0.8\, -\, 1.2]$ where a change in the phase implies a deviation of the on-axis intensity by more than 5 $\%$ in linear propagation regime \cite{Ouadghiri2017}.

\begin{figure}[htb]
\centering
\includegraphics[width=0.48\textwidth]{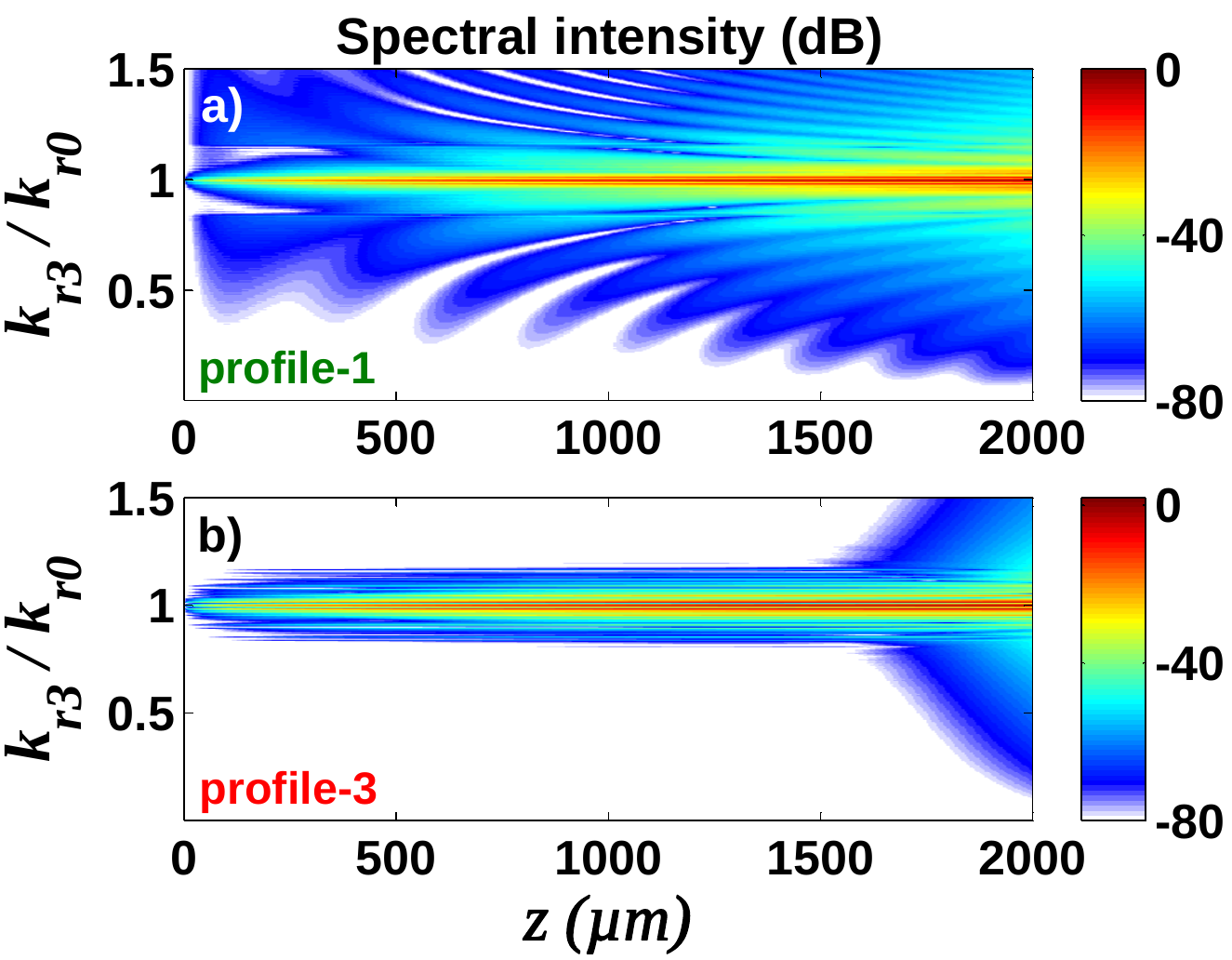}
\caption{\label{BG_Iz2_FWM} Results using our FWM model: frequency resolved intensity (dB) of the signal wave along propagation resulting from the interaction of three pump waves composed each of twenty-seven frequencies: $k_r/k_{r0} = [ 0.805\ -\ 1.195]$ for input phase distributions of (a) profile-1 and (b) profile-3.}
\end{figure}

 As in previous section, we will interpret the reduced nonlinear instabilities for profile 3 by the phase mismatch in the very broad pump spectrum.
 Using our reduced FWM model described by Eq. (\ref{da3_reduced}), we consider the contribution of 27 spectral components in the chosen spectral range. This number has simply been chosen  as the minimum relevant number of waves necessary to obtain a quantitative agreement with NLSE simulation results.  We numerically compute the spectral intensity of the  wave using Eq. (\ref{da3_reduced})  and compare its evolution for spatial spectra corresponding to profiles 1 and 3. The input spectral phases are the same as in Fig. \ref{BB_shape_sim}(d) and the relative amplitudes of the components are also described by the respective spatial spectra of the profiles. 
 
 Our results are shown in Fig. \ref{BG_Iz2_FWM}. Profile 1 (BG beam) shows the same parabolic structures as previously. In contrast, this feature is totally absent in the case of profile 3 where spectral broadening only occurs starting from a propagation distance of $z\approx 1700$ $\mu$m. Particularly, the spectral expansion is in good qualitative agreement with NLSE simulation results of Fig. \ref{BB_shape_sim}(c). We then conclude that the phase of the low intensity tails of the spatial spectrum also contribute to the initial spectral broadening stage.

 We stress that these conclusions are theoretically only valid in case the phase distribution is preserved along propagation. According to our numerical simulation of the NLSE, the relative input phase remains approximately unchanged up to a propagation point of $z = 1200~\mu$m. As the spectral phase flattens because of linear propagation \cite{Jarutis2000}, FWM interactions become more efficient and inevitably lead to significant growth of the axial wave and outer ring. 
 
 % 

%Although our approach allows a decrease in nonlinear oscillations, complete suppression of these oscillations remains a difficult challenge{, especially regarding the optimal input spectral intensity of the additional wave around $k_r/k_{r0} = 1.5$ and the corresponding spectral phase profile}. Further theoretical studies are required to complete this approach. \ 

%%%%%%%%%%%%%%%%%%%%%%%%%%%%%%%%%%%%%%%%%%%%%%%%%%%%%%%%%%%%%%%%%%%%%%%%%%%%%%%%%%%%%%%%%%%%%%%%%%%%%%%%%%%%%%%%%%%%%
\section{Conclusions}
%\addcontentsline{toc}{section}{Conclusions}

In conclusion, we have developed a Four Wave Mixing model in order to interpret the different characteristics of the growth of Kerr-induced instabilities in Bessel beams.  We have shown that nonlinear spectral distortions are established in two steps. The first step consists of spectral broadening and the generation of an axial wave seed. In a second step, this axial wave seed is amplified  and an outer ring is generated via Four Wave Mixing. These new spatial frequency components interfere with the main input Bessel beam, yielding oscillations of the on-axis intensity. 

We have then used a reduced model where only the dominant wave mixing terms were present so that we could expand the model to broad spatial spectra. This reduced nonlinear model allowed us to understand previous experimental and numerical results where input conditions of Bessel beams were yielding very different growth rates of nonlinear instabilities. We have demonstrated that these differences can be well explained from the weak differences in input spectral phases, even outside the main peak of the spectral amplitude.  We believe this approach will open new ways to  control nonlinear instabilities, as well as to extend the applicability of Bessel beams for new applications. We also note that further work is needed to expand the model to ultrashort pulses with broadband temporal frequency spectrum. 

The research leading to these results has received funding from the European Research Council (ERC) under the European Union's Horizon 2020 research and innovation program (grant agreement No 682032-PULSAR). This work has been supported by the EIPHI Graduate School (contract ANR-17-EURE-0002).


%apsrev4-2.bst 2018-12-27 (MD) hand-edited version of apsrev4-1.bst
%Control: key (0)
%Control: author (8) initials jnrlst
%Control: editor formatted (1) identically to author
%Control: production of article title (0) allowed
%Control: page (0) single
%Control: year (1) truncated
%Control: production of eprint (0) enabled
\begin{thebibliography}{26}%
\makeatletter
\providecommand \@ifxundefined [1]{%
 \@ifx{#1\undefined}
}%
\providecommand \@ifnum [1]{%
 \ifnum #1\expandafter \@firstoftwo
 \else \expandafter \@secondoftwo
 \fi
}%
\providecommand \@ifx [1]{%
 \ifx #1\expandafter \@firstoftwo
 \else \expandafter \@secondoftwo
 \fi
}%
\providecommand \natexlab [1]{#1}%
\providecommand \enquote  [1]{``#1''}%
\providecommand \bibnamefont  [1]{#1}%
\providecommand \bibfnamefont [1]{#1}%
\providecommand \citenamefont [1]{#1}%
\providecommand \href@noop [0]{\@secondoftwo}%
\providecommand \href [0]{\begingroup \@sanitize@url \@href}%
\providecommand \@href[1]{\@@startlink{#1}\@@href}%
\providecommand \@@href[1]{\endgroup#1\@@endlink}%
\providecommand \@sanitize@url [0]{\catcode `\\12\catcode `\$12\catcode
  `\&12\catcode `\#12\catcode `\^12\catcode `\_12\catcode `\%12\relax}%
\providecommand \@@startlink[1]{}%
\providecommand \@@endlink[0]{}%
\providecommand \url  [0]{\begingroup\@sanitize@url \@url }%
\providecommand \@url [1]{\endgroup\@href {#1}{\urlprefix }}%
\providecommand \urlprefix  [0]{URL }%
\providecommand \Eprint [0]{\href }%
\providecommand \doibase [0]{https://doi.org/}%
\providecommand \selectlanguage [0]{\@gobble}%
\providecommand \bibinfo  [0]{\@secondoftwo}%
\providecommand \bibfield  [0]{\@secondoftwo}%
\providecommand \translation [1]{[#1]}%
\providecommand \BibitemOpen [0]{}%
\providecommand \bibitemStop [0]{}%
\providecommand \bibitemNoStop [0]{.\EOS\space}%
\providecommand \EOS [0]{\spacefactor3000\relax}%
\providecommand \BibitemShut  [1]{\csname bibitem#1\endcsname}%
\let\auto@bib@innerbib\@empty
%</preamble>
\bibitem [{\citenamefont {Durnin}(1987)}]{Durnin1987}%
  \BibitemOpen
  \bibfield  {author} {\bibinfo {author} {\bibfnamefont {J.}~\bibnamefont
  {Durnin}},\ }\bibfield  {title} {\bibinfo {title} {Exact solutions for
  nondiffracting beams i the scalar theory},\ }\href
  {https://doi.org/10.1364/josaa.4.000651} {\bibfield  {journal} {\bibinfo
  {journal} {Journal of the Optical Society of America A}\ }\textbf {\bibinfo
  {volume} {4}},\ \bibinfo {pages} {651} (\bibinfo {year} {1987})}\BibitemShut
  {NoStop}%
\bibitem [{\citenamefont {Porras}\ \emph {et~al.}(2004)\citenamefont {Porras},
  \citenamefont {Parola}, \citenamefont {Faccio}, \citenamefont {Dubietis},\
  and\ \citenamefont {DiTrapani}}]{Porras2004}%
  \BibitemOpen
  \bibfield  {author} {\bibinfo {author} {\bibfnamefont {M.~A.}\ \bibnamefont
  {Porras}}, \bibinfo {author} {\bibfnamefont {A.}~\bibnamefont {Parola}},
  \bibinfo {author} {\bibfnamefont {D.}~\bibnamefont {Faccio}}, \bibinfo
  {author} {\bibfnamefont {A.}~\bibnamefont {Dubietis}}, and\ \bibinfo {author}
  {\bibfnamefont {P.}~\bibnamefont {DiTrapani}},\ }\bibfield  {title} {\bibinfo
  {title} {Nonlinear unbalanced {Bessel} beams: Stationary conical waves
  supported by nonlinear losses},\ }\href
  {https://doi.org/10.1103/physrevlett.93.153902} {\bibfield  {journal}
  {\bibinfo  {journal} {Physical Review Letters}\ }\textbf {\bibinfo {volume}
  {93}},\ \bibinfo {pages} {153902} (\bibinfo {year} {2004})}\BibitemShut
  {NoStop}%
\bibitem [{\citenamefont {Polesana}\ \emph {et~al.}(2008)\citenamefont
  {Polesana}, \citenamefont {Franco}, \citenamefont {Couairon}, \citenamefont
  {Faccio},\ and\ \citenamefont {DiTrapani}}]{Polesana2008}%
  \BibitemOpen
  \bibfield  {author} {\bibinfo {author} {\bibfnamefont {P.}~\bibnamefont
  {Polesana}}, \bibinfo {author} {\bibfnamefont {M.}~\bibnamefont {Franco}},
  \bibinfo {author} {\bibfnamefont {A.}~\bibnamefont {Couairon}}, \bibinfo
  {author} {\bibfnamefont {D.}~\bibnamefont {Faccio}}, and\ \bibinfo {author}
  {\bibfnamefont {P.}~\bibnamefont {DiTrapani}},\ }\bibfield  {title} {\bibinfo
  {title} {Filamentation in kerr media from pulsed {Bessel} beams},\ }\href
  {https://doi.org/10.1103/physreva.77.043814} {\bibfield  {journal} {\bibinfo
  {journal} {Physical Review A}\ }\textbf {\bibinfo {volume} {77}},\ \bibinfo
  {pages} {043814} (\bibinfo {year} {2008})}\BibitemShut {NoStop}%
\bibitem [{\citenamefont {Dota}\ \emph {et~al.}(2012)\citenamefont {Dota},
  \citenamefont {Pathak}, \citenamefont {Dharmadhikari}, \citenamefont
  {Mathur},\ and\ \citenamefont {Dharmadhikari}}]{Dota2012}%
  \BibitemOpen
  \bibfield  {author} {\bibinfo {author} {\bibfnamefont {K.}~\bibnamefont
  {Dota}}, \bibinfo {author} {\bibfnamefont {A.}~\bibnamefont {Pathak}},
  \bibinfo {author} {\bibfnamefont {J.~A.}\ \bibnamefont {Dharmadhikari}},
  \bibinfo {author} {\bibfnamefont {D.}~\bibnamefont {Mathur}}, and\ \bibinfo
  {author} {\bibfnamefont {A.~K.}\ \bibnamefont {Dharmadhikari}},\ }\bibfield
  {title} {\bibinfo {title} {Femtosecond laser filamentation in condensed media
  with {Bessel} beams},\ }\href {https://doi.org/10.1103/physreva.86.023808}
  {\bibfield  {journal} {\bibinfo  {journal} {Physical Review A}\ }\textbf
  {\bibinfo {volume} {86}},\ \bibinfo {pages} {023808} (\bibinfo {year}
  {2012})}\BibitemShut {NoStop}%
\bibitem [{\citenamefont {Bhuyan}\ \emph {et~al.}(2010)\citenamefont {Bhuyan},
  \citenamefont {Courvoisier}, \citenamefont {Lacourt}, \citenamefont
  {Jacquot}, \citenamefont {Salut}, \citenamefont {Furfaro},\ and\
  \citenamefont {Dudley}}]{Bhuyan2010}%
  \BibitemOpen
  \bibfield  {author} {\bibinfo {author} {\bibfnamefont {M.~K.}\ \bibnamefont
  {Bhuyan}}, \bibinfo {author} {\bibfnamefont {F.}~\bibnamefont {Courvoisier}},
  \bibinfo {author} {\bibfnamefont {P.~A.}\ \bibnamefont {Lacourt}}, \bibinfo
  {author} {\bibfnamefont {M.}~\bibnamefont {Jacquot}}, \bibinfo {author}
  {\bibfnamefont {R.}~\bibnamefont {Salut}}, \bibinfo {author} {\bibfnamefont
  {L.}~\bibnamefont {Furfaro}}, and\ \bibinfo {author} {\bibfnamefont {J.~M.}\
  \bibnamefont {Dudley}},\ }\bibfield  {title} {\bibinfo {title} {High aspect
  ratio nanochannel machining using single shot femtosecond {Bessel} beams},\
  }\href {https://doi.org/10.1063/1.3479419} {\bibfield  {journal} {\bibinfo
  {journal} {Applied Physics Letters}\ }\textbf {\bibinfo {volume} {97}},\
  \bibinfo {pages} {081102} (\bibinfo {year} {2010})}\BibitemShut {NoStop}%
\bibitem [{\citenamefont {Bhuyan}\ \emph {et~al.}(2011)\citenamefont {Bhuyan},
  \citenamefont {Courvoisier}, \citenamefont {Phing}, \citenamefont
  {Jedrkiewicz}, \citenamefont {Recchia}, \citenamefont {Trapani},\ and\
  \citenamefont {Dudley}}]{Bhuyan2011}%
  \BibitemOpen
  \bibfield  {author} {\bibinfo {author} {\bibfnamefont {M.~K.}\ \bibnamefont
  {Bhuyan}}, \bibinfo {author} {\bibfnamefont {F.}~\bibnamefont {Courvoisier}},
  \bibinfo {author} {\bibfnamefont {H.~S.}\ \bibnamefont {Phing}}, \bibinfo
  {author} {\bibfnamefont {O.}~\bibnamefont {Jedrkiewicz}}, \bibinfo {author}
  {\bibfnamefont {S.}~\bibnamefont {Recchia}}, \bibinfo {author} {\bibfnamefont
  {P.~D.}\ \bibnamefont {Trapani}}, and\ \bibinfo {author} {\bibfnamefont
  {J.~M.}\ \bibnamefont {Dudley}},\ }\bibfield  {title} {\bibinfo {title}
  {Laser micro- and nanostructuring using femtosecond {Bessel} beams},\ }\href
  {https://doi.org/10.1140/epjst/e2011-01506-0} {\bibfield  {journal} {\bibinfo
   {journal} {The European Physical Journal Special Topics}\ }\textbf {\bibinfo
  {volume} {199}},\ \bibinfo {pages} {101} (\bibinfo {year}
  {2011})}\BibitemShut {NoStop}%
\bibitem [{\citenamefont {Garzillo}\ \emph {et~al.}(2016)\citenamefont
  {Garzillo}, \citenamefont {Jukna}, \citenamefont {Couairon}, \citenamefont
  {Grigutis}, \citenamefont {Trapani},\ and\ \citenamefont
  {Jedrkiewicz}}]{Garzillo2016}%
  \BibitemOpen
  \bibfield  {author} {\bibinfo {author} {\bibfnamefont {V.}~\bibnamefont
  {Garzillo}}, \bibinfo {author} {\bibfnamefont {V.}~\bibnamefont {Jukna}},
  \bibinfo {author} {\bibfnamefont {A.}~\bibnamefont {Couairon}}, \bibinfo
  {author} {\bibfnamefont {R.}~\bibnamefont {Grigutis}}, \bibinfo {author}
  {\bibfnamefont {P.~D.}\ \bibnamefont {Trapani}}, and\ \bibinfo {author}
  {\bibfnamefont {O.}~\bibnamefont {Jedrkiewicz}},\ }\bibfield  {title}
  {\bibinfo {title} {Optimization of laser energy deposition for single-shot
  high aspect-ratio microstructuring of thick {BK}7 glass},\ }\href
  {https://doi.org/10.1063/1.4954890} {\bibfield  {journal} {\bibinfo
  {journal} {Journal of Applied Physics}\ }\textbf {\bibinfo {volume} {120}},\
  \bibinfo {pages} {013102} (\bibinfo {year} {2016})}\BibitemShut {NoStop}%
\bibitem [{\citenamefont {Courvoisier}\ \emph {et~al.}(2016)\citenamefont
  {Courvoisier}, \citenamefont {Stoian},\ and\ \citenamefont
  {Couairon}}]{Courvoisier2016}%
  \BibitemOpen
  \bibfield  {author} {\bibinfo {author} {\bibfnamefont {F.}~\bibnamefont
  {Courvoisier}}, \bibinfo {author} {\bibfnamefont {R.}~\bibnamefont {Stoian}},
  and\ \bibinfo {author} {\bibfnamefont {A.}~\bibnamefont {Couairon}},\
  }\bibfield  {title} {\bibinfo {title} {[{INVITED}] ultrafast laser micro- and
  nano-processing with nondiffracting and curved beams},\ }\href
  {https://doi.org/10.1016/j.optlastec.2015.11.026} {\bibfield  {journal}
  {\bibinfo  {journal} {Optics {\&} Laser Technology}\ }\textbf {\bibinfo
  {volume} {80}},\ \bibinfo {pages} {125} (\bibinfo {year} {2016})}\BibitemShut
  {NoStop}%
\bibitem [{\citenamefont {Stoian}\ \emph {et~al.}(2018)\citenamefont {Stoian},
  \citenamefont {Bhuyan}, \citenamefont {Zhang}, \citenamefont {Cheng},
  \citenamefont {Meyer},\ and\ \citenamefont {Courvoisier}}]{Stoian2018}%
  \BibitemOpen
  \bibfield  {author} {\bibinfo {author} {\bibfnamefont {R.}~\bibnamefont
  {Stoian}}, \bibinfo {author} {\bibfnamefont {M.~K.}\ \bibnamefont {Bhuyan}},
  \bibinfo {author} {\bibfnamefont {G.}~\bibnamefont {Zhang}}, \bibinfo
  {author} {\bibfnamefont {G.}~\bibnamefont {Cheng}}, \bibinfo {author}
  {\bibfnamefont {R.}~\bibnamefont {Meyer}}, and\ \bibinfo {author}
  {\bibfnamefont {F.}~\bibnamefont {Courvoisier}},\ }\bibfield  {title}
  {\bibinfo {title} {Ultrafast {Bessel} beams: advanced tools for laser
  materials processing},\ }\href {https://doi.org/10.1515/aot-2018-0009}
  {\bibfield  {journal} {\bibinfo  {journal} {Advanced Optical Technologies}\
  }\textbf {\bibinfo {volume} {7}},\ \bibinfo {pages} {165} (\bibinfo {year}
  {2018})}\BibitemShut {NoStop}%
\bibitem [{\citenamefont {Xie}\ \emph {et~al.}(2015)\citenamefont {Xie},
  \citenamefont {Jukna}, \citenamefont {Mili{\'{a}}n}, \citenamefont {Giust},
  \citenamefont {Ouadghiri-Idrissi}, \citenamefont {Itina}, \citenamefont
  {Dudley}, \citenamefont {Couairon},\ and\ \citenamefont
  {Courvoisier}}]{Xie2015}%
  \BibitemOpen
  \bibfield  {author} {\bibinfo {author} {\bibfnamefont {C.}~\bibnamefont
  {Xie}}, \bibinfo {author} {\bibfnamefont {V.}~\bibnamefont {Jukna}}, \bibinfo
  {author} {\bibfnamefont {C.}~\bibnamefont {Mili{\'{a}}n}}, \bibinfo {author}
  {\bibfnamefont {R.}~\bibnamefont {Giust}}, \bibinfo {author} {\bibfnamefont
  {I.}~\bibnamefont {Ouadghiri-Idrissi}}, \bibinfo {author} {\bibfnamefont
  {T.}~\bibnamefont {Itina}}, \bibinfo {author} {\bibfnamefont {J.~M.}\
  \bibnamefont {Dudley}}, \bibinfo {author} {\bibfnamefont {A.}~\bibnamefont
  {Couairon}}, and\ \bibinfo {author} {\bibfnamefont {F.}~\bibnamefont
  {Courvoisier}},\ }\bibfield  {title} {\bibinfo {title} {Tubular filamentation
  for laser material processing},\ }\bibfield  {journal} {\bibinfo  {journal}
  {Scientific Reports}\ }\textbf {\bibinfo {volume} {5}},\ \href
  {https://doi.org/10.1038/srep08914} {10.1038/srep08914} (\bibinfo {year}
  {2015})\BibitemShut {NoStop}%
\bibitem [{\citenamefont {Gadonas}\ \emph {et~al.}(2001)\citenamefont
  {Gadonas}, \citenamefont {Jarutis}, \citenamefont {Pa{\v{s}}kauskas},
  \citenamefont {Smilgevi{\v{c}}ius}, \citenamefont {Stabinis},\ and\
  \citenamefont {Vai{\v{c}}aitis}}]{Gadonas2001}%
  \BibitemOpen
  \bibfield  {author} {\bibinfo {author} {\bibfnamefont {R.}~\bibnamefont
  {Gadonas}}, \bibinfo {author} {\bibfnamefont {V.}~\bibnamefont {Jarutis}},
  \bibinfo {author} {\bibfnamefont {R.}~\bibnamefont {Pa{\v{s}}kauskas}},
  \bibinfo {author} {\bibfnamefont {V.}~\bibnamefont {Smilgevi{\v{c}}ius}},
  \bibinfo {author} {\bibfnamefont {A.}~\bibnamefont {Stabinis}}, and\ \bibinfo
  {author} {\bibfnamefont {V.}~\bibnamefont {Vai{\v{c}}aitis}},\ }\bibfield
  {title} {\bibinfo {title} {Self-action of {Bessel} beam in nonlinear
  medium},\ }\href {https://doi.org/10.1016/s0030-4018(01)01386-4} {\bibfield
  {journal} {\bibinfo  {journal} {Optics Communications}\ }\textbf {\bibinfo
  {volume} {196}},\ \bibinfo {pages} {309} (\bibinfo {year}
  {2001})}\BibitemShut {NoStop}%
\bibitem [{\citenamefont {Andreev}\ \emph {et~al.}(1991)\citenamefont
  {Andreev}, \citenamefont {Aristov}, \citenamefont {Polonskii},\ and\
  \citenamefont {Pyatnitskii}}]{Andreev1991}%
  \BibitemOpen
  \bibfield  {author} {\bibinfo {author} {\bibfnamefont {N.~E.}\ \bibnamefont
  {Andreev}}, \bibinfo {author} {\bibfnamefont {Y.~A.}\ \bibnamefont
  {Aristov}}, \bibinfo {author} {\bibfnamefont {L.~Y.}\ \bibnamefont
  {Polonskii}}, and\ \bibinfo {author} {\bibfnamefont {L.~N.}\ \bibnamefont
  {Pyatnitskii}},\ }\bibfield  {title} {\bibinfo {title} {Bessel beams of
  electromagnetic waves: self-effect and nonlinear structures},\ }\href@noop {}
  {\bibfield  {journal} {\bibinfo  {journal} {Zhurnal Eksperimentalnoi i
  Teoreticheskoi Fiziki}\ }\textbf {\bibinfo {volume} {100}},\ \bibinfo {pages}
  {1756} (\bibinfo {year} {1991})}\BibitemShut {NoStop}%
\bibitem [{\citenamefont {Pyragaite}\ \emph {et~al.}(2006)\citenamefont
  {Pyragaite}, \citenamefont {Regelskis}, \citenamefont {Smilgevicius},\ and\
  \citenamefont {Stabinis}}]{Pyragaite2006}%
  \BibitemOpen
  \bibfield  {author} {\bibinfo {author} {\bibfnamefont {V.}~\bibnamefont
  {Pyragaite}}, \bibinfo {author} {\bibfnamefont {K.}~\bibnamefont
  {Regelskis}}, \bibinfo {author} {\bibfnamefont {V.}~\bibnamefont
  {Smilgevicius}}, and\ \bibinfo {author} {\bibfnamefont {A.}~\bibnamefont
  {Stabinis}},\ }\bibfield  {title} {\bibinfo {title} {Self-action of {Bessel}
  light beams in medium with large nonlinearity},\ }\href
  {https://doi.org/10.1016/j.optcom.2005.07.012} {\bibfield  {journal}
  {\bibinfo  {journal} {Optics Communications}\ }\textbf {\bibinfo {volume}
  {257}},\ \bibinfo {pages} {139} (\bibinfo {year} {2006})}\BibitemShut
  {NoStop}%
\bibitem [{\citenamefont {Polesana}\ \emph {et~al.}(2007)\citenamefont
  {Polesana}, \citenamefont {Couairon}, \citenamefont {Faccio}, \citenamefont
  {Parola}, \citenamefont {Porras}, \citenamefont {Dubietis}, \citenamefont
  {Piskarskas},\ and\ \citenamefont {DiTrapani}}]{Polesana2007}%
  \BibitemOpen
  \bibfield  {author} {\bibinfo {author} {\bibfnamefont {P.}~\bibnamefont
  {Polesana}}, \bibinfo {author} {\bibfnamefont {A.}~\bibnamefont {Couairon}},
  \bibinfo {author} {\bibfnamefont {D.}~\bibnamefont {Faccio}}, \bibinfo
  {author} {\bibfnamefont {A.}~\bibnamefont {Parola}}, \bibinfo {author}
  {\bibfnamefont {M.~A.}\ \bibnamefont {Porras}}, \bibinfo {author}
  {\bibfnamefont {A.}~\bibnamefont {Dubietis}}, \bibinfo {author}
  {\bibfnamefont {A.}~\bibnamefont {Piskarskas}}, and\ \bibinfo {author}
  {\bibfnamefont {P.}~\bibnamefont {DiTrapani}},\ }\bibfield  {title} {\bibinfo
  {title} {Observation of conical waves in focusing, dispersive, and
  dissipative {Kerr} media},\ }\href
  {https://doi.org/10.1103/physrevlett.99.223902} {\bibfield  {journal}
  {\bibinfo  {journal} {Physical Review Letters}\ }\textbf {\bibinfo {volume}
  {99}},\ \bibinfo {pages} {223902} (\bibinfo {year} {2007})}\BibitemShut
  {NoStop}%
\bibitem [{\citenamefont {Johannisson}\ \emph {et~al.}(2003)\citenamefont
  {Johannisson}, \citenamefont {Anderson}, \citenamefont {Lisak},\ and\
  \citenamefont {Marklund}}]{Johannisson2003}%
  \BibitemOpen
  \bibfield  {author} {\bibinfo {author} {\bibfnamefont {P.}~\bibnamefont
  {Johannisson}}, \bibinfo {author} {\bibfnamefont {D.}~\bibnamefont
  {Anderson}}, \bibinfo {author} {\bibfnamefont {M.}~\bibnamefont {Lisak}},
  and\ \bibinfo {author} {\bibfnamefont {M.}~\bibnamefont {Marklund}},\
  }\bibfield  {title} {\bibinfo {title} {Nonlinear {Bessel} beams},\ }\href
  {https://doi.org/10.1016/s0030-4018(03)01603-1} {\bibfield  {journal}
  {\bibinfo  {journal} {Optics Communications}\ }\textbf {\bibinfo {volume}
  {222}},\ \bibinfo {pages} {107} (\bibinfo {year} {2003})}\BibitemShut
  {NoStop}%
\bibitem [{\citenamefont {Couairon}\ \emph {et~al.}(2013)\citenamefont
  {Couairon}, \citenamefont {Lotti}, \citenamefont {Panagiotopoulos},
  \citenamefont {Abdollahpour}, \citenamefont {Faccio}, \citenamefont
  {Papazoglou}, \citenamefont {Tzortzakis}, \citenamefont {Courvoisier},\ and\
  \citenamefont {Dudley}}]{Couairon2013}%
  \BibitemOpen
  \bibfield  {author} {\bibinfo {author} {\bibfnamefont {A.}~\bibnamefont
  {Couairon}}, \bibinfo {author} {\bibfnamefont {A.}~\bibnamefont {Lotti}},
  \bibinfo {author} {\bibfnamefont {P.}~\bibnamefont {Panagiotopoulos}},
  \bibinfo {author} {\bibfnamefont {D.}~\bibnamefont {Abdollahpour}}, \bibinfo
  {author} {\bibfnamefont {D.}~\bibnamefont {Faccio}}, \bibinfo {author}
  {\bibfnamefont {D.~G.}\ \bibnamefont {Papazoglou}}, \bibinfo {author}
  {\bibfnamefont {S.}~\bibnamefont {Tzortzakis}}, \bibinfo {author}
  {\bibfnamefont {F.}~\bibnamefont {Courvoisier}}, and\ \bibinfo {author}
  {\bibfnamefont {J.~M.}\ \bibnamefont {Dudley}},\ }\bibfield  {title}
  {\bibinfo {title} {Ultrashort laser pulse filamentation with {Airy} and
  {Bessel} beams},\ }in\ \href {https://doi.org/10.1117/12.2014198} {\emph
  {\bibinfo {booktitle} {17th International School on Quantum Electronics:
  Laser Physics and Applications}}},\ \bibinfo {editor} {edited by\ \bibinfo
  {editor} {\bibfnamefont {T.~N.}\ \bibnamefont {Dreischuh}}and\ \bibinfo
  {editor} {\bibfnamefont {A.~T.}\ \bibnamefont {Daskalova}}}\ (\bibinfo
  {publisher} {{SPIE}},\ \bibinfo {year} {2013})\BibitemShut {NoStop}%
\bibitem [{\citenamefont {Gaizauskas}\ \emph {et~al.}(2006)\citenamefont
  {Gaizauskas}, \citenamefont {Vanagas}, \citenamefont {Jarutis}, \citenamefont
  {Juodkazis}, \citenamefont {Mizeikis},\ and\ \citenamefont
  {Misawa}}]{Gaizauskas2006}%
  \BibitemOpen
  \bibfield  {author} {\bibinfo {author} {\bibfnamefont {E.}~\bibnamefont
  {Gaizauskas}}, \bibinfo {author} {\bibfnamefont {E.}~\bibnamefont {Vanagas}},
  \bibinfo {author} {\bibfnamefont {V.}~\bibnamefont {Jarutis}}, \bibinfo
  {author} {\bibfnamefont {S.}~\bibnamefont {Juodkazis}}, \bibinfo {author}
  {\bibfnamefont {V.}~\bibnamefont {Mizeikis}}, and\ \bibinfo {author}
  {\bibfnamefont {H.}~\bibnamefont {Misawa}},\ }\bibfield  {title} {\bibinfo
  {title} {Discrete damage traces from filamentation of {Gauss-Bessel}
  pulses},\ }\href {https://doi.org/10.1364/ol.31.000080} {\bibfield  {journal}
  {\bibinfo  {journal} {Optics Letters}\ }\textbf {\bibinfo {volume} {31}},\
  \bibinfo {pages} {80} (\bibinfo {year} {2006})}\BibitemShut {NoStop}%
\bibitem [{\citenamefont {Ouadghiri-Idrissi}\ \emph {et~al.}(2017)\citenamefont
  {Ouadghiri-Idrissi}, \citenamefont {Dudley},\ and\ \citenamefont
  {Courvoisier}}]{Ouadghiri2017}%
  \BibitemOpen
  \bibfield  {author} {\bibinfo {author} {\bibfnamefont {I.}~\bibnamefont
  {Ouadghiri-Idrissi}}, \bibinfo {author} {\bibfnamefont {J.~M.}\ \bibnamefont
  {Dudley}}, and\ \bibinfo {author} {\bibfnamefont {F.}~\bibnamefont
  {Courvoisier}},\ }\bibfield  {title} {\bibinfo {title} {Controlling nonlinear
  instabilities in {Bessel} beams through longitudinal intensity shaping},\
  }\href {https://doi.org/10.1364/ol.42.003785} {\bibfield  {journal} {\bibinfo
   {journal} {Optics Letters}\ }\textbf {\bibinfo {volume} {42}},\ \bibinfo
  {pages} {3785} (\bibinfo {year} {2017})}\BibitemShut {NoStop}%
\bibitem [{\citenamefont {Porras}\ \emph {et~al.}(2015)\citenamefont {Porras},
  \citenamefont {Ruiz-Jim{\'{e}}nez},\ and\ \citenamefont
  {Losada}}]{Porras2015}%
  \BibitemOpen
  \bibfield  {author} {\bibinfo {author} {\bibfnamefont {M.~A.}\ \bibnamefont
  {Porras}}, \bibinfo {author} {\bibfnamefont {C.}~\bibnamefont
  {Ruiz-Jim{\'{e}}nez}}, and\ \bibinfo {author} {\bibfnamefont {J.~C.}\
  \bibnamefont {Losada}},\ }\bibfield  {title} {\bibinfo {title} {Underlying
  conservation and stability laws in nonlinear propagation of axicon-generated
  bessel beams},\ }\href {https://doi.org/10.1103/physreva.92.063826}
  {\bibfield  {journal} {\bibinfo  {journal} {Physical Review A}\ }\textbf
  {\bibinfo {volume} {92}},\ \bibinfo {pages} {063826} (\bibinfo {year}
  {2015})}\BibitemShut {NoStop}%
\bibitem [{\citenamefont {Porras}\ \emph {et~al.}(2016)\citenamefont {Porras},
  \citenamefont {Carvalho}, \citenamefont {Leblond},\ and\ \citenamefont
  {Malomed}}]{Porras2016}%
  \BibitemOpen
  \bibfield  {author} {\bibinfo {author} {\bibfnamefont {M.~A.}\ \bibnamefont
  {Porras}}, \bibinfo {author} {\bibfnamefont {M.}~\bibnamefont {Carvalho}},
  \bibinfo {author} {\bibfnamefont {H.}~\bibnamefont {Leblond}}, and\ \bibinfo
  {author} {\bibfnamefont {B.~A.}\ \bibnamefont {Malomed}},\ }\bibfield
  {title} {\bibinfo {title} {Stabilization of vortex beams in {Kerr} media by
  nonlinear absorption},\ }\href {https://doi.org/10.1103/physreva.94.053810}
  {\bibfield  {journal} {\bibinfo  {journal} {Physical Review A}\ }\textbf
  {\bibinfo {volume} {94}},\ \bibinfo {pages} {053810} (\bibinfo {year}
  {2016})}\BibitemShut {NoStop}%
\bibitem [{\citenamefont {Tewari}\ \emph {et~al.}(1996)\citenamefont {Tewari},
  \citenamefont {Huang},\ and\ \citenamefont {Boyd}}]{Tewari1996}%
  \BibitemOpen
  \bibfield  {author} {\bibinfo {author} {\bibfnamefont {S.~P.}\ \bibnamefont
  {Tewari}}, \bibinfo {author} {\bibfnamefont {H.}~\bibnamefont {Huang}}, and\
  \bibinfo {author} {\bibfnamefont {R.~W.}\ \bibnamefont {Boyd}},\ }\bibfield
  {title} {\bibinfo {title} {Theory of third-harmonic generation using {Bessel}
  beams, and self-phase-matching},\ }\href
  {https://doi.org/10.1103/physreva.54.2314} {\bibfield  {journal} {\bibinfo
  {journal} {Physical Review A}\ }\textbf {\bibinfo {volume} {54}},\ \bibinfo
  {pages} {2314} (\bibinfo {year} {1996})}\BibitemShut {NoStop}%
\bibitem [{DLM(2018)}]{DLMF}%
  \BibitemOpen
  \href {http://dlmf.nist.gov} {\bibinfo {title} {{NIST} digital library of
  mathematical functions}},\ \bibinfo {howpublished} {http://dlmf.nist.gov/,
  Release 1.0.19 of 2018-06-22} (\bibinfo {year} {2018}),\ \bibinfo {note}
  {f.~W.~J. Olver, A.~B. {Olde Daalhuis}, D.~W. Lozier, B.~I. Schneider, R.~F.
  Boisvert, C.~W. Clark, B.~R. Miller and B.~V. Saunders, eds.}\BibitemShut
  {Stop}%
\bibitem [{\citenamefont {Couairon}\ \emph {et~al.}(2011)\citenamefont
  {Couairon}, \citenamefont {Brambilla}, \citenamefont {Corti}, \citenamefont
  {Majus}, \citenamefont {de~J.~Ram{\'{\i}}rez-G{\'{o}}ngora},\ and\
  \citenamefont {Kolesik}}]{Couairon2011}%
  \BibitemOpen
  \bibfield  {author} {\bibinfo {author} {\bibfnamefont {A.}~\bibnamefont
  {Couairon}}, \bibinfo {author} {\bibfnamefont {E.}~\bibnamefont {Brambilla}},
  \bibinfo {author} {\bibfnamefont {T.}~\bibnamefont {Corti}}, \bibinfo
  {author} {\bibfnamefont {D.}~\bibnamefont {Majus}}, \bibinfo {author}
  {\bibfnamefont {O.}~\bibnamefont {de~J.~Ram{\'{\i}}rez-G{\'{o}}ngora}}, and\
  \bibinfo {author} {\bibfnamefont {M.}~\bibnamefont {Kolesik}},\ }\bibfield
  {title} {\bibinfo {title} {Practitioner's guide to laser pulse propagation
  models and simulation},\ }\href {https://doi.org/10.1140/epjst/e2011-01503-3}
  {\bibfield  {journal} {\bibinfo  {journal} {The European Physical Journal
  Special Topics}\ }\textbf {\bibinfo {volume} {199}},\ \bibinfo {pages} {5}
  (\bibinfo {year} {2011})}\BibitemShut {NoStop}%
\bibitem [{\citenamefont {Jarutis}\ \emph {et~al.}(2000)\citenamefont
  {Jarutis}, \citenamefont {Pa{\v{s}}kauskas},\ and\ \citenamefont
  {Stabinis}}]{Jarutis2000}%
  \BibitemOpen
  \bibfield  {author} {\bibinfo {author} {\bibfnamefont {V.}~\bibnamefont
  {Jarutis}}, \bibinfo {author} {\bibfnamefont {R.}~\bibnamefont
  {Pa{\v{s}}kauskas}}, and\ \bibinfo {author} {\bibfnamefont {A.}~\bibnamefont
  {Stabinis}},\ }\bibfield  {title} {\bibinfo {title} {Focusing of
  {Laguerre{\textendash}Gaussian} beams by axicon},\ }\href
  {https://doi.org/10.1016/s0030-4018(00)00961-5} {\bibfield  {journal}
  {\bibinfo  {journal} {Optics Communications}\ }\textbf {\bibinfo {volume}
  {184}},\ \bibinfo {pages} {105} (\bibinfo {year} {2000})}\BibitemShut
  {NoStop}%
\bibitem [{\citenamefont {Sogomonian}\ \emph {et~al.}(1999)\citenamefont
  {Sogomonian}, \citenamefont {Barille},\ and\ \citenamefont
  {Rivoire}}]{Sogomonian1999}%
  \BibitemOpen
  \bibfield  {author} {\bibinfo {author} {\bibfnamefont {S.}~\bibnamefont
  {Sogomonian}}, \bibinfo {author} {\bibfnamefont {R.}~\bibnamefont {Barille}},
  and\ \bibinfo {author} {\bibfnamefont {G.}~\bibnamefont {Rivoire}},\
  }\bibfield  {title} {\bibinfo {title} {Spatial distortions of a {Bessel} beam
  in a {Kerr}-type medium},\ }in\ \href {https://doi.org/10.1117/12.375305}
  {\emph {\bibinfo {booktitle} {New Trends in Atomic and Molecular
  Spectroscopy}}},\ \bibinfo {editor} {edited by\ \bibinfo {editor}
  {\bibfnamefont {G.~G.}\ \bibnamefont {Gurzadyan}}and\ \bibinfo {editor}
  {\bibfnamefont {A.~V.}\ \bibnamefont {Karmenyan}}}\ (\bibinfo  {publisher}
  {{SPIE}},\ \bibinfo {year} {1999})\BibitemShut {NoStop}%
\bibitem [{\citenamefont {Dubietis}\ \emph {et~al.}(2007)\citenamefont
  {Dubietis}, \citenamefont {Polesana}, \citenamefont {Valiulis}, \citenamefont
  {Stabinis}, \citenamefont {Trapani},\ and\ \citenamefont
  {Piskarskas}}]{Dubietis2007}%
  \BibitemOpen
  \bibfield  {author} {\bibinfo {author} {\bibfnamefont {A.}~\bibnamefont
  {Dubietis}}, \bibinfo {author} {\bibfnamefont {P.}~\bibnamefont {Polesana}},
  \bibinfo {author} {\bibfnamefont {G.}~\bibnamefont {Valiulis}}, \bibinfo
  {author} {\bibfnamefont {A.}~\bibnamefont {Stabinis}}, \bibinfo {author}
  {\bibfnamefont {P.~D.}\ \bibnamefont {Trapani}}, and\ \bibinfo {author}
  {\bibfnamefont {A.}~\bibnamefont {Piskarskas}},\ }\bibfield  {title}
  {\bibinfo {title} {Axial emission and spectral broadening in self-focusing of
  femtosecond {Bessel} beams},\ }\href {https://doi.org/10.1364/oe.15.004168}
  {\bibfield  {journal} {\bibinfo  {journal} {Optics Express}\ }\textbf
  {\bibinfo {volume} {15}},\ \bibinfo {pages} {4168} (\bibinfo {year}
  {2007})}\BibitemShut {NoStop}%
\end{thebibliography}%


%apsrev4-2.bst 2018-12-27 (MD) hand-edited version of apsrev4-1.bst
%Control: key (0)
%Control: author (72) initials jnrlst
%Control: editor formatted (1) identically to author
%Control: production of article title (-1) disabled
%Control: page (0) single
%Control: year (1) truncated
%Control: production of eprint (0) enabled
\begin{thebibliography}{0}%
\makeatletter
\providecommand \@ifxundefined [1]{%
 \@ifx{#1\undefined}
}%
\providecommand \@ifnum [1]{%
 \ifnum #1\expandafter \@firstoftwo
 \else \expandafter \@secondoftwo
 \fi
}%
\providecommand \@ifx [1]{%
 \ifx #1\expandafter \@firstoftwo
 \else \expandafter \@secondoftwo
 \fi
}%
\providecommand \natexlab [1]{#1}%
\providecommand \enquote  [1]{``#1''}%
\providecommand \bibnamefont  [1]{#1}%
\providecommand \bibfnamefont [1]{#1}%
\providecommand \citenamefont [1]{#1}%
\providecommand \href@noop [0]{\@secondoftwo}%
\providecommand \href [0]{\begingroup \@sanitize@url \@href}%
\providecommand \@href[1]{\@@startlink{#1}\@@href}%
\providecommand \@@href[1]{\endgroup#1\@@endlink}%
\providecommand \@sanitize@url [0]{\catcode `\\12\catcode `\$12\catcode
  `\&12\catcode `\#12\catcode `\^12\catcode `\_12\catcode `\%12\relax}%
\providecommand \@@startlink[1]{}%
\providecommand \@@endlink[0]{}%
\providecommand \url  [0]{\begingroup\@sanitize@url \@url }%
\providecommand \@url [1]{\endgroup\@href {#1}{\urlprefix }}%
\providecommand \urlprefix  [0]{URL }%
\providecommand \Eprint [0]{\href }%
\providecommand \doibase [0]{https://doi.org/}%
\providecommand \selectlanguage [0]{\@gobble}%
\providecommand \bibinfo  [0]{\@secondoftwo}%
\providecommand \bibfield  [0]{\@secondoftwo}%
\providecommand \translation [1]{[#1]}%
\providecommand \BibitemOpen [0]{}%
\providecommand \bibitemStop [0]{}%
\providecommand \bibitemNoStop [0]{.\EOS\space}%
\providecommand \EOS [0]{\spacefactor3000\relax}%
\providecommand \BibitemShut  [1]{\csname bibitem#1\endcsname}%
\let\auto@bib@innerbib\@empty
%</preamble>
\end{thebibliography}%
\end{document}